\documentclass[prd,aps,a4paper,twocolumn,nofootinbib,superscriptaddress]{revtex4-2}

\usepackage[utf8]{inputenc}
\usepackage[normalem]{ulem}

\usepackage{graphicx,psfrag}
\usepackage{mathrsfs}
\usepackage{amsmath,amsfonts,amssymb}
\usepackage{multirow}
\usepackage{comment}
\usepackage{bm}
\usepackage{epstopdf}
\usepackage{mathrsfs}
\usepackage{comment}
\usepackage{multirow}
\usepackage{hyperref}
\usepackage{diagbox} 
\usepackage{amssymb,amsthm,wasysym}
\usepackage{capt-of}
\usepackage{physics}
\usepackage[dvipsnames]{xcolor}
\usepackage{orcidlink}
\usepackage{placeins}

\definecolor{darkblue}{rgb}{0.0, 0.1, 0.7}
\hypersetup{
    colorlinks=true,
    linkcolor=darkblue,
    filecolor=darkblue,      
    urlcolor=darkblue,
    citecolor = darkblue,
    pdftitle={Spherical inspirals of spinning bodies into Kerr black holes},
    }

\renewcommand{\textbf}{}

\DeclareMathOperator{\sgn}{sgn}
\newcommand{\wmean}{\vec{\mathsf{w}}}

\begin{document}

\title{Spherical inspirals of spinning bodies into Kerr black holes} 

\author{Viktor Skoup\'y \orcidlink{0000-0001-7475-5324}}
\email{viktor.skoupy@matfyz.cuni.cz}

\affiliation{Institute of Theoretical Physics, Faculty of Mathematics and Physics, Charles University, CZ-180 00 Prague, Czech Republic}

\author{Gabriel Andres Piovano \orcidlink{0000-0003-1782-6813}}
\email{gabriel.andres.piovano@ulb.be}

\affiliation{Université Libre de Bruxelles, BLU-ULB Brussels Laboratory of the Universe, International
Solvay Institutes, CP 231, B-1050 Brussels, Belgium}
\affiliation{School of Mathematics and Statistics, University College Dublin, Belfield, Dublin 4, Ireland}

\author{Vojtěch Witzany \orcidlink{0000-0002-9209-5355}}
\email{vojtech.witzany@matfyz.cuni.cz}

\affiliation{Institute of Theoretical Physics, Faculty of Mathematics and Physics, Charles University, CZ-180 00 Prague, Czech Republic}

\begin{abstract}

Extreme mass-ratio inspirals (EMRIs), consisting of a stellar-mass compact object spiraling into a massive black hole, are key sources for future space-based gravitational wave observatories such as LISA. Accurate modeling of these systems requires incorporating the spin effects of both the primary and secondary bodies, particularly for waveforms at the precision required for LISA detection and astrophysical parameter extraction. In this work, we develop a framework for modeling flux-driven spherical inspirals (orbits of approximately constant Boyer-Lindquist radius) of a spinning secondary into a Kerr black hole. We leverage recently found solutions for the motion of spinning test particles and compute the associated gravitational wave fluxes to linear order in the secondary spin. Next, we show that spherical orbits remain spherical under radiation reaction at linear order in spin, and derive the evolution of the orbital parameters throughout the inspiral. We implement a numerical scheme for waveform generation in the frequency domain and assess the impact of the secondary spin on the gravitational wave signal. In contrast to quasi-circular inspirals, we find that neglecting the secondary spin in spherical inspirals induces large mismatches in the waveforms that will plausibly be detectable by LISA.

\end{abstract}

\maketitle

\tableofcontents

\section{Introduction}\label{sec:intro}
Upcoming space-based gravitational wave (GW) observatories such as LISA \cite{LISA:2017pwj,LISA:2024hlh}, TianQin \cite{TianQin:2015yph}, and Taiji \cite{Ruan:2020smc} will be sensitive to extreme mass-ratio inspirals (EMRIs).  These systems consist of a stellar-mass compact object spiraling into a massive black hole (MBH). With mass ratios $q = \mu/M \sim 10^{-4} - 10^{-7}$, EMRIs will complete $\sim 10^4 - 10^6$ orbital cycles in the detector bands, providing unprecedented opportunities to probe the strong-field regime of general relativity and measure MBH masses and spins with high precision \cite{Barack:2006pq,Gair:2012nm, Babak:2017tow}.

The detections amassed by the LIGO-Virgo-KAGRA collaboration to date correspond to binaries with small mass asymmetries and no discernible non-circularity. However, next-generation GW detectors such as Einstein Telescope or Cosmic Explorer promise to detect a broader spectrum of signals from various formation channels \cite{Reitze:2019iox,Abac:2025saz}. Among these channels, we expect to see some fraction of dynamically formed binaries with nonzero eccentricities, larger mass asymmetries, and misaligned spins and angular momenta \cite{Mapelli:2021gyv}.

To fully exploit these scientific opportunities, accurate theoretical models must account for all significant physical effects. The accurate modeling of EMRIs and binaries with large mass asymmetries in general requires accounting for the spins of both the heavy primaries and the lighter secondaries. The primary's spin, when misaligned with the orbital angular momentum, causes precession of the orbital plane of the motion. On the other hand, the secondary's spin introduces small oscillations of the orbital plane and other subtle effects \cite{Witzany:2023bmq,Skoupy:2024uan}. In the self-force approach \cite{Poisson:2011nh, Barack:2018yvs, Pound:2021}, EMRIs are described iteratively starting from a motion of a test particle in the space-time of the primary MBH field. In this framework, the primary spin contribution appears as the spin in the Kerr metric representing the MBH field, while the contribution of the secondary spin is included through the dynamics of a spinning test particle in curved space-time, which is governed by the Mathisson-Papapetrou-Dixon (MPD) equations \cite{Mathisson:2010, Papapetrou:1951pa, Dixon:1970I}.

More specifically, the contributions needed to accurately model EMRIs are best described within a two-timescale expansion that exploits the separation between radiation reaction and orbital motion timescales \cite{Hinderer:2008dm,Pound:2021}. In short, one systematically takes into account that the orbital decay due to radiation of GWs is much slower than the orbital motion. The phase elapsed during inspiral can then be decomposed as,
\begin{align}
    \phi_{\rm insp} = \frac{1}{q} \phi_{\rm 0 PA} + \phi_{\rm 1 PA} + \mathcal{O}(q)\,,
\end{align}
where $\phi_{n\mathrm{PA}}$ are $\mathcal{O}(1)$ coefficients. The leading adiabatic order (0PA) derives from point-mass geodesic motion and associated GW fluxes \cite{Kennefick:1995za,Hughes:2021}. The post-adiabatic (1PA) order includes several critical contributions, notably the secondary object's spin effects \cite{Mathews:2025}. By Lindblom's criterion \cite{Lindblom:2008cm}, modeling to 1PA accuracy ensures waveforms indistinguishable from true signals for signal to noise ratios (SNRs) up to $\mathcal{O}(1/q)$, while neglecting 1PA contributions produces potentially distinguishable errors even at modest SNRs \cite{Burke:2023lno}. Therefore, computing the secondary spin contribution is essential for accurate EMRI modeling. Additionally, the decomposition provides valuable information for comparable mass ratio formalisms. The spin contributions to the 1PA phase can be obtained from two ingredients: the motion of spinning test particles in Kerr space-time to linear order in the secondary spin, and the linear correction to GW fluxes that is introduced by the secondary spin along the orbit. \textbf{Is this model sufficient to capture the 1PA effects of secondary spin, or should one perhaps include quadratic and higher spin orders? The specific spin magnitude $s = S/\mu$ in the equations of motion is of order $\order{q}$ in the large mass ratio limit~\cite{Hartl:2002ig,Piovano:2020zin}. The $\order{S^2}$ corrections can then be shown to contribute only to 2PA and higher order terms~\cite{Mathews:2025}.}

In this paper, we focus on a specific but important class of EMRI systems: quasi-spherical inspirals. These are orbits that maintain an approximately constant Boyer-Lindquist radius while allowing arbitrary inclination with respect to the primary's spin axis. Additionally, we allow an arbitrarily inclined secondary spin. These orbits represent an important class of EMRI systems for several reasons: they can arise through astrophysical formation channels like tidal separation of comparable mass binaries by a nearby MBH (the Hills mechanism) \cite{ColemanMiller:2005rm}; radiation reaction tends to circularize eccentric orbits \cite{peters1964gravitational,Glampedakis:2002cb}; and they serve as a crucial stepping stone toward understanding generic EMRI dynamics. Specifically, while the eccentricity is efficiently damped already at leading order in the weak field, the spin-orbital inclination does not evolve until the 1.5 post-Newtonian order, and one thus generally expects binaries to enter detector bands at inclinations close to the ones with which they were formed \cite{Rodriguez:2016vmx,Farr:2017uvj}. In the case of strong-field EMRIs, Kennefick and Ori \cite{Kennefick:1995za} showed that spherical orbits of non-spinning particles in Kerr space-time remain spherical under adiabatic radiation reaction. This implies that spherical orbits in the sense of a constant Boyer-Lindquist radius are the end state of eccentricity relaxation.

The understanding of the motion of spinning test bodies in Kerr spacetime has seen major progress in recent years. At linear order in spin, sufficient constants of motion exist to achieve integrability \cite{Rudiger:1981, Rudiger:1983}. The Hamilton-Jacobi equation for the spinning-particle Hamiltonian has been solved perturbatively, reducing the equations of motion to first order \cite{Witzany:2018ahb, Witzany:2019}. Solutions for the fundamental frequencies have been derived analytically \cite{Witzany:2024ttz}, and various numerical and semi-analytical approaches for computing spinning particle trajectories have been developed \cite{Drummond:2022efc, Drummond:2022xej, Piovano:2024}. Recently, a closed-form analytical solution for the motion to linear order in the secondary spin was found \cite{Skoupy:2024uan}.

GW fluxes generated by spinning particles, the second ingredient needed for the 1PA contribution to the phase, have also been tackled recently. Fluxes from spinning particles on circular orbits were calculated in Refs.~\cite{Harms:2016ctx,Akcay:2019bvk} for Schwarzschild spacetime, and in Refs.~\cite{Tanaka:1996ht,Han:2010tp,Harms:2015ixa,Lukes-Gerakopoulos:2017vkj,Piovano:2020zin} for Kerr spacetime. This calculation was extended to eccentric orbits in both Schwarzschild and Kerr spacetimes in Refs.~\cite{Skoupy:2021asz,Skoupy:2022adh,Skoupy:2024jsi}. In Kerr spacetime, the orientation of orbits relative to the Kerr spin axis affects the resulting fluxes. The fluxes for spinning particles on both eccentric and inclined orbits were recently computed by~\citet{Skoupy:2023} and~\citet{Piovano:2024}.

Our work extends and connects these investigations by providing a comprehensive framework for modeling the flux-driven spherical inspirals of spinning bodies into Kerr black holes,
and the corresponding waveforms.%
\footnote{Formally, the evolution is an \textit{adiabatic} inspiral of a spinning particle, since its orbital decay is only driven by flux-balance laws. However, we use this to compute a contribution to the first \textit{post-adiabatic} term in the phase. We thus use the term \textit{flux-driven} to avoid confusion. }
 In other words: we compute fully precessing waveforms of spinning compact binary inspirals in the large mass ratio limit. 

We leverage recent developments in the analytical understanding of spinning particle motion and combine them with GW flux calculations to derive consistent equations of motion for the inspiral.
 Our approach accounts for the coupling between the secondary's spin, the primary's spin, and the radiation reaction forces. As a result, we obtain the full 1PA contribution to the phase induced by secondary spin in spherical inspirals, thus providing a foundational piece for accurate EMRI waveform modeling. We also use these results to provide a preliminary assessment of the detectability of secondary spin in EMRIs by computing waveform mismatches. Our results suggest that secondary spins should be detectable by LISA in spherical inspirals. 

The paper is organized as follows. In Sec. \ref{sec:spinorb}, we introduce our semi-analytical descriptions of nearly spherical trajectories of spinning test particles in Kerr space-time. We start by introducing spherical geodesics, continue by discussing parallel transport and, finally, we present the general solutions for spherical orbits of spinning particles and discuss their parametrization. Section \ref{sec:GW_fluxes} focuses on the calculation of GW fluxes from spinning bodies on spherical orbits. Next, in Sec. \ref{sec:inspirals} we first prove that spherical orbits remain spherical under dissipation through fluxes to linear order in spin and show how to use the fluxes to drive the inspiral. In Sec. \ref{sec:numerical_implementation} we describe our numerical approach to model the inspirals, calculate their waveforms and compare waveforms with spinning and non-spinning secondary. Finally, we conclude in Sec. \ref{sec:Concl} with a discussion of our results and directions for future work.

\subsection{Notation} 
Throughout this paper, we use geometrized units, where $G = c = 1$, and the $(-,+,+,+)$ signature of the metric. Spacetime indices are denoted by lowercase Greek letters, and the covariant derivative is denoted by a semicolon. The Riemann tensor is defined as $a_{\nu;\kappa\lambda} - a_{\nu;\lambda\kappa} \equiv R^\mu{}_{\nu\kappa\lambda} a_\mu$, and for the Levi-Civita tensor in $t,r,z,\phi$ coordinates we use the sign convention given by $\epsilon^{trz\phi}\sqrt{-g} = - \epsilon_{trz\phi}/\sqrt{-g} = -1$.

\section{Spinning bodies orbiting a Kerr black hole}
\label{sec:spinorb}

A spinning test body in Kerr spacetime experiences both gravitational and spin-orbit coupling effects. The motion remains integrable at linear order in the secondary spin, allowing us to construct analytical solutions that capture the essential physics while remaining computationally tractable. In this Section, we describe the motion of spinning bodies in the Kerr space-time in the specific case of nearly spherical orbits. 

In Boyer-Lindquist-like coordinates $(t,r,z=\cos\theta,\phi)$ the metric tensor reads
\begin{multline}
    \dd s^2 = -\qty( 1 - \frac{2 M r}{\Sigma} ) \dd t^2 - \frac{4 a M r (1-z^2)}{\Sigma} \dd t \dd \phi\, + \\ \frac{\qty(\varpi^4 - a^2 \Delta (1-z^2))(1-z^2)}{\Sigma} \dd \phi^2 + \frac{\Sigma}{\Delta} \dd r^2 + \frac{\Sigma}{1-z^2} \dd z^2 \,,
\end{multline}
where
\begin{align*}
    \Sigma &= r^2 + a^2 z^2 \; , \\
    \Delta &= r^2 - 2 M r + a^2 \; , \\
    \varpi^2 &= r^2+a^2 \; .
\end{align*}

In the pole-dipole approximation, the motion of compact spinning bodies in curved spacetime is governed by the so-called Mathisson-Papapetrou-Dixon equations \cite{Mathisson:2010, Papapetrou:1951pa, Dixon:1970I}
\begin{align}
    \frac{D P^\mu}{\dd \tau} &= - \frac{1}{2} R^\mu{}_{\nu\kappa\lambda} \dv{x^\nu}{\tau} S^{\kappa\lambda} \,, \label{eq:Pmpd}\\
    \frac{D S^{\mu\nu}}{\dd \tau} &= P^{\mu} \dv{x^{\nu}}{\tau} - P^{\nu} \dv{x^{\mu}}{\tau} \,, \label{eq:Smpd}
\end{align}
where $P^\mu$ is the four-momentum, $\tau$ is the proper time, $R^{\mu}{}_{\nu\kappa\lambda}$ is the Riemann tensor, $x^\mu(\tau)$ 
is the representative worldline from within the body, and $S^{\mu\nu} = -S^{\nu\mu}$ is the spin tensor. In particular, the momentum-evolution equation \eqref{eq:Pmpd} contains the spin-curvature term that relays both the spin-orbital and spin-spin effects to the orbital degrees of freedom. To fix the center of mass, we use the Tulczyjew-Dixon spin-supplementary condition (SSC) $S^{\mu\nu} P_\nu = 0$ in this work \cite{tulczyjew1959motion,Dixon:1970I}. Under this SSC, the magnitude of the spin $S$ and the mass $\mu$ defined as
\begin{equation}
    S = \sqrt{\frac{1}{2} S^{\mu\nu} S_{\mu\nu}} \, , \qquad \mu = \sqrt{- P^\mu P_\mu}
\end{equation}
are conserved.

We can define specific four-momentum and specific spin tensor as
\begin{equation}
    u^\mu = \frac{P^\mu}{\mu} \, , \qquad s^{\mu\nu} = \frac{S^{\mu\nu}}{\mu} \,,
\end{equation}
and analogously the specific spin magnitude as $s = S/\mu$.

Due to the presence of Killing vectors $\xi_{(t)}^\mu \partial_\mu = \partial_t$ and $\xi_{(\phi)}^\mu \partial_\mu = \partial_\phi$, there exist two constants of motion in the form \cite{Dixon:1970I}
\begin{align}
    E = - u_\mu \xi_{(t)}^\mu + \frac{1}{2} \xi^{(t)}_{\mu;\nu} s^{\mu\nu} \,, \\
    L_z = u_\mu \xi_{(\phi)}^\mu - \frac{1}{2} \xi^{(\phi)}_{\mu;\nu} s^{\mu\nu} \,,
\end{align}
which can be interpreted as the total specific energy and the component of the total specific angular momentum parallel to the symmetry axis. 

It is easy to see that in geometric units, the spin magnitude $S$ has the dimension of the angular momentum, thus
\begin{equation}
\frac{S}{\mu M} = q \chi \,,
\end{equation}
where $q =\mu/M$ is the small mass ratio, and $\chi = S/\mu^2$ is the dimensionless spin. For the EMRI binaries detectable by LISA, $\chi \leq 1$.%
\footnote{An EMRI is observable by LISA if the mass of the central black hole  is within $10^{5.5} -10^7 M_\odot$. In this mass range, the MBH Hill mass is such that main-sequence stars or brown dwarfs are tidally disrupted before reaching the last stable orbit~\cite{Gezari:2014td}. 
White dwarfs, neutron stars and black holes are the only known astrophysical objects that are compact enough to plunge into the central black hole for these EMRIs. Black holes have spin $\chi \leq 1$ due to the Kerr bound, while typical neutron stars have $\chi \sim \mathcal (1/10)$ and $\chi\lesssim 0.6$ due to the mass-shedding limit~\cite{Hartl:2002ig}. White dwarfs can have spins $\chi \sim 10$, but the product $q \chi \ll 1$ for EMRIs.} 

As described in \cite{Skoupy:2023}, it is sufficient for EMRIs to truncate the equations of motion at the linear-in-spin order $\mathcal O(q \chi)$. Then, the four-momentum and four-velocity are collinear ($P^\mu = \mu \dv*{\tilde{z}^\mu}{\tau}$) and the particle's trajectory can be expanded as $\tilde z = x_{\rm g} + q  \chi \delta x$, with $x^\mu_{\rm g}$ a referential geodesic, and $\delta x$ the correction due to the test body's spin. The  equations of motion then take the form
\begin{align}
    \frac{D^2 x^\mu_{\rm g}}{\dd \tau^2} &= 0 \,, \\
    \frac{D^2 \delta x^\mu}{\dd \tau^2} &= - \frac{1}{2} R^\mu{}_{\nu\kappa\lambda} \dv{x^\nu_{\rm g}}{\tau} s^{\kappa\lambda} \,, \\
    \frac{D s^{\mu\nu}}{\dd \tau} &= 0 \,.
\end{align}
Due to the Killing-Yano tensor $Y_{\mu\nu}$, $Y_{\mu(\nu;\kappa)} = 0$, in the linearized regime, there are two additional constants of motion \cite{Rudiger:1981,Rudiger:1983}
\begin{align}
    s_\parallel &= \frac{Y_{\mu\nu} u^\mu s^\nu}{\sqrt{K_{\mu\nu} u^\mu u^\nu}} \,, \\
    K &= K_{\mu\nu} u^\mu u^\nu + 4 u^\mu s^{\rho\sigma} Y^\kappa{}_{\left[\mu\right.} Y_{\left.\sigma\right]\rho;\kappa} \,,
\end{align}
where we defined a specific spin vector
\begin{equation}
    s^\mu = - \frac{1}{2} \epsilon^{\mu\nu\rho\sigma} u_\nu s_{\rho\sigma} \,.
\end{equation}
The constants $s_\parallel$ and $K$ are conserved up to linear order in spin, i.e. 
\begin{equation}
    \dv{s_\parallel}{\tau} = \order{S^2} = \dv{K}{\tau} \,.
\end{equation}
These constants of motion can be interpreted as components of the spin vector parallel to the orbital angular momentum and analog to the Carter constant. Moreover, we separate the spin magnitude $\chi$ into two components: $\chi_\parallel = s_\parallel/\mu$, which is parallel to the orbital angular momentum, and  $\chi_\perp = \sqrt{\chi^2 - \chi_\parallel^2}$. Likewise, we define $s_\perp = \sqrt{s^2 - s_\parallel^2} = \mu \chi_\perp$.

In general, the constants of motion $E, L_z$ and $K$ depend on both $\chi_\parallel$ and $\chi_\perp$  whereas their time-averages only depend on $\chi_\parallel$ at first order in $\mathcal O(q \chi)$~\cite{Witzany:2019,Drummond:2022efc,Drummond:2022xej,Piovano:2024,Skoupy:2024uan}. We can therefore choose a referential worldline close to the spinning test-body worldline by writing
\begin{align}
E &= E_{\rm g} + q \chi_\parallel \delta E \, , \\
L_{z} &= L_{z\rm g} + q \chi_\parallel \delta L_z \, , \\
K &= K_{z\rm g} + q \chi_\parallel \delta K \, , 
\end{align}
with $E_{\rm g}$, $L_{z\rm g}$ and $K_{z\rm g}$ the constants of motion of a geodesic trajectory, and $\delta E$, $\delta L_z$ and $\delta K$ the corresponding averaged corrections due to the small body spin
(see Refs.~\cite{Piovano:2024,Skoupy:2024uan} for generic bound orbits). 
We will use the array $(C_{1 \rm g}, C_{2 \rm g}, C_{3 \rm g}) =(E_{\rm g}, L_{z \rm g}, K_{\rm g})$ to represent the geodesic constant of motion while $\delta C_i$ will denote the corresponding shifts induced by the spinning particle, that is, $ (\delta C_1, \delta C_2, \delta C_3) = (\delta E, \delta L_z, \delta K)$. 
Section~\ref{ref:fix_the_gauge} discusses the spin-corrections to the constants of motion in a specific gauge.

\subsection{Spherical geodesics in Kerr spacetime} \label{ref:geodesic_motion}

Spherical orbits represent the end state of eccentricity evolution under radiation reaction, making them natural building blocks for EMRI modeling. Here, we briefly describe spherical geodesics in Kerr space-time, a base ingredient to which spin effects will later be added. A more detailed account can be found in~\cite{Schmidt:2002, Teo:2021}. 

The first-order geodesic equations become fully separable when parametrized by Mino time $\dd \lambda = \dd \tau/\Sigma$. Under this parametrization they are given by
\begin{align}
    \dv{t_{\rm g}}{\lambda} &= V^t_{r\text{g}}(r_{\rm g}) + V^t_{z\text{g}}(z_{\rm g}) \, , \\
    \qty(\dv{r_{\rm g}}{\lambda})^{\!2} &= R_\text{g}(r_{\rm g}) \, , \\
    \qty(\dv{z_{\rm g}}{\lambda})^{\!2} &= Z_\text{g}(z_{\rm g}) \, , \\
    \dv{\phi_{\rm g}}{\lambda} &=  V^\phi_{r\text{g}}(r_{\rm g}) + V^\phi_{z\text{g}}(z_{\rm g}) \, , 
\end{align}
where
\begin{align}
& V^t_{r \rm g}(r_{\rm g}) = E_{\rm g} \frac{(r_{\rm g}^2 + a^2)^2}{\Delta} - 2 \frac{a\, r_{\rm g}}{\Delta} L_{z \rm g}  \, ,\\
& V^t_{z\rm g}(z_{\rm g}) =  - a^2 E_{\rm g}  (1-z_{\rm g}^2) \, ,\\
& R_{\rm g}(r_{\rm g}) = P_r(r_{\rm g})^2 - \Delta (K_{\rm g} + r_{\rm g}^2 )  \, , \\
& Z_{\rm g} (z_{\rm g}) = - P_z(z_{\rm g})^2 + (1-z^2_{\rm g})(K_{\rm g}-a^2 z^2_{\rm g}) \, , \\
& V^\phi_{r \rm g}(r_{\rm g})  = \frac{a}{\Delta} P_r(r_{\rm g}) -a E_{\rm g}  \, , \\
& V^\phi_{z \rm g}(z_{\rm g})  = \frac{L_{z \rm g}}{1-z_{\rm g}^2}  \,,
\end{align}
and we have further introduced 
\begin{align}
    P_r(r_{\rm g}) &= E_{\rm g}(r^2_{\rm g}+a^2)- aL_{z \rm g} \, , \\
    P_z(z_{\rm g}) &=  L_{z\rm g} - (1-z^2_{\rm g}) a E_{\rm g} \, .
\end{align}

It is common in the literature to parametrize the radial geodesic motion in Kerr space time using a Kepler-like parametrization
\begin{equation}\label{eq:r_param}
r_{\rm g} = \frac{p}{1+ e \cos(\chi_r)} \,,
\end{equation}
where $p$ is a semi-latus rectum and $e$ an eccentricity parameter, while $\chi_r$ is the relativistic anomaly~\cite{darwin1959gravity,Schmidt:2002}. For a spherical orbit $e=0$ thus $p = r_{\rm g}$. Geodesic spherical orbits satisfy the constraint
\begin{subequations}\label{eq:conditions_turning_points_geodesic}
\begin{align} 
    R_\text{g}(r_\text{g}) &= 0 \,,\\
    R_\text{g}'(r_\text{g}) &= 0 \, . 
\end{align}
\end{subequations}
\textbf{Here and in what follows, we use the prime to denote a derivative of a function of a single argument with respect to that argument.}

Since the function $R_{\rm g}(r_{\rm g})$ is a quartic polynomial in $r_{\rm g}$, the condition for spherical orbits implies that $R_{\rm g}$ has a double root, which is the radial position of the particle. The remaining radial roots $r_{3\mathrm{g}}$ and $r_{4\mathrm{g}}$ are given in~\cite{Fujita:2009bp} and satisfy the condition $r_{4\mathrm{g}} < r_{3\mathrm{g}} \leq r_{\mathrm{g}}$. 

The polar potential $Z_{\rm g}(z_{\rm g})$ is also a quartic polynomial in $z_{\rm g}$, which can be conveniently factorized as~\cite{Schmidt:2002}
\begin{align}
&Z_{\rm g}(z_{\rm g}) = (z_{1\rm g}^2 - z^2_{\rm g}) Y_{z\rm g}^2(z_{\rm g})\, ,
\end{align}
where 
\begin{align}
&Y_{z\rm g}(z_{\rm g}) = \sqrt{z^2_{2 \rm g} - a^2(1-E^2_{\rm g})z^2_{\rm g}} \, . 
\end{align}
Bound motion is confined within $-z_{ 1 \rm g}\leq z_{\rm g} \leq z_{ 1 \rm g}$, with the root $z_{2\rm g}$ given by
\begin{equation}
    z_{2\mathrm{g}}=\sqrt{a^2(1-E_{\mathrm{g}}^2)+\frac{L_{\mathrm{g}}}{1-z_{1\mathrm{g}}}}.
\end{equation}
The polar motion can be parametrized as
\begin{equation}
z_{\rm g} = \sqrt{1 - x^2_{\rm g}} \sin(\chi_z)
\end{equation}
with $x_{\rm g}$ a parameter describing the inclination of the orbit with respect to the equatorial plane, defined as
\begin{equation}
x_{\rm g} = \sgn(L_{z\rm g}) \sqrt{1 -z^2_{1 \rm g}} = \sgn(L_{z\rm g}) \cos\theta_\text{min}.
\end{equation}
and $z_{1 \rm g} \in [0,1]$ the physical root of the polar potential. Thus, $x_{\rm g} \in [-1,1]$, with positive (negative) values corresponds to prograde (retrograde) orbits. Moreover, $x_{\rm g} = \pm 1$ indicates prograde or retrograde equatorial orbits. 

The motion is periodic in Mino time with polar frequency $\Upsilon_{z\text{g}}$, azimuthal frequency $\Upsilon_{\phi\text{g}}$ and average rate of change (``frequency'') of the coordinate time $\Upsilon_{t\text{g}}$. The coordinate time frequencies are then calculated as
\begin{equation}
    \Omega_{z\text{g}} = \frac{\Upsilon_{z\text{g}}}{\Upsilon_{t\text{g}}} \,, \qquad \Omega_{\phi\text{g}} = \frac{\Upsilon_{\phi\text{g}}}{\Upsilon_{t\text{g}}} \,.
\end{equation}

Analytic solutions for geodesic spherical orbits can be inferred from the generic bound motion expressions given in Refs.~\cite{Fujita:2009bp,vandeMeent:2020}. Throughout the paper, we assume that the fiducial geodesic, which describes the motion at zeroth order in the spin, is a spherical orbit.

\subsection{Solution for the spin vector}

The evolution of the spin vector in the linear regime can be expressed as parallel transport along the trajectory
\begin{equation}
    \frac{D s^\mu}{\dd \tau} = 0 + \mathcal{O}(q^2 \chi^2)\,.
\end{equation}
Since the spin effects are suppressed by one power of the mass ratio, the parallel transport can be calculated along a geodesic. This problem can be solved using the Marck tetrad $u^\mu_{\rm g}$, $e_1^\mu$, $e_2^\mu$, $e_3^\mu$ \cite{Marck:1983}, which is parallel transported along the reference geodesic. Then, the spin vector reads
\begin{equation}
    s^\mu = s_\parallel e_3^\mu + s_\perp \qty(e_1^\mu \cos\psi + e_2^\mu \sin\psi) \,,
\end{equation}
where $\psi(\tau)$ is the precession phase. Conveniently, the precession phase of the Marck tetrad evolves according to the separable equation  
\begin{equation}
    \dv{\psi}{\lambda} = \Psi_r(r_{\rm g}) +  \Psi_z(z_{\rm g})  \, , \label{eq:spin-precession-angle}
\end{equation}
where
\begin{align}
\Psi_r(r_{\rm g}) &= \sqrt{K_{\rm g}}\frac{(r^2_{\rm g}+a^2)E_{\rm g}-a L_{z \rm g}}{K_\text{g} + r^2_{\rm g}}  \, , \\
\Psi_z(z_{\rm g}) &= a\sqrt{K_{\rm g}}\frac{L_{z \rm g} - a (1-z^2_{\rm g})E_{\rm g}}{K_{\rm g} - a^2 z^2_{\rm g}} \, .
\end{align}
Here we follow the convention for the Marck tetrad given by~\cite{vandeMeent:2020}  (see~\cite{Skoupy:2024uan} for other conventions). The average rate of change of $\psi$, the precession frequency, is $\Upsilon_\text{p}$. See \cite{vandeMeent:2020} for the analytic solution for $\psi(\lambda)$ and $\Upsilon_\text{p}$.

\subsection{Nearly spherical orbits for a spinning particle} \label{sec:general_solutions_spherical_orbits}

Building on these foundational results for geodesics, we now turn to spin corrections to spherical orbits. In what follows, we will use the term ``nearly spherical'' to denote orbits of spinning particles that have a spherical geodesic as their referential geodesic. This implies that they must be spherical up to $\mathcal{O}(q \chi)$ corrections. We will see that the spin corrections will not make it possible to keep the spinning-particle orbits exactly spherical, and we will thus also need to fix a class of nearly spherical orbits useful for our context. 

\subsubsection{Radial motion}

The radial motion of a nearly spherical orbit is given entirely by spin corrections. It is described by Eq.~(46a) of Ref.~\cite{Witzany:2019},
\begin{equation}
\qty( \frac{\dd r}{\dd \lambda} )^{\! 2} = R_g(r_{\rm g}) + q \delta R(r_{\rm g}, z_{\rm g}, \psi)  + \mathcal O(q^2 \chi^2) \,, \label{eq:radial_motion}
\end{equation}
where
\begin{align}
\delta R &= R_{\rm s} + R'_{\rm g}(r_{\rm g})\delta r + \displaystyle \sum^{3}_{i=1}\frac{\partial R_{\rm g}}{\partial C_{i\rm g}} \delta C_i \,, \label{eq:diff_radial_potential} \\
R_{\rm s} &= R^{\text{sep}}_{\rm s}(r_{\rm g}) + 2 \Delta^2 \mathcal S^r \,, \\
R^{\text{sep}}_{\rm s}(r_{\rm g}) &= 2 \Delta \big( \Psi_{r}(r_{\rm g}) + \delta K_{\rm{eq}} \big) \,,
\end{align}
with $\mathcal S^r = \mathcal S^r(r_{\rm g},z_{\rm g}, \psi)$ one of the spin-connection terms (see Appendix C of~\cite{Piovano:2024}), $\delta K_{\rm{eq}} = a \chi_\parallel \text{sgn}(L_{z\rm g} - a E_{\rm g})$, while a prime denotes derivatives with respect to $r_{\rm g}$. The radial shift $\delta r = \delta r(r_{\rm g}, z_{\rm g},\psi)$ is given as
\begin{align}
\delta r &= \chi_\parallel \delta r_\parallel(r_{\rm g}, z_{\rm g}) + \chi_\perp\delta r_\perp(r_{\rm g}, z_{\rm g},\psi) .
\end{align}
Let us now examine the possibility of spherical orbits of spinning test particles defined by the constraints
\begin{align}
R_{\rm g}(r_{\rm g}) &= 0 = R'_{\rm g}(r_{\rm g}) \,, \\
\delta R(r_{\rm g},z_{\rm g},\psi) &= 0 = \frac{\partial}{\partial r_{\rm g}}\delta R(r_{\rm g},z_{\rm g},\psi) \,, \end{align}
which imply
\begin{align}
&R^{\text{sep}}_{\rm s}(r_{\rm g})  + \displaystyle \sum^{3}_{i=1}\frac{\partial R_{\rm g}}{\partial C_{i\rm g}} \delta C_i =0 \,, \label{eq:constraint_spherical_correction} \\
&\frac{\partial}{\partial r_{\rm g}}R_{\rm s}(r_{\rm g},z_{\rm g},\psi) -2 Y^2_{r\rm g}(r_{\rm g})\delta r +  \displaystyle \sum^{3}_{i=1}\frac{\partial R'_{\rm g}(r_{\rm g})}{\partial C_{i\rm g}} \delta C_i =0 \,,\label{eq:constraint_spherical_correction-derivative}
\end{align}
where $Y_{r\rm g}(r_{\rm g}) = \sqrt{(1 - E^2_{\rm g})(r_{\rm g} - r_{3\rm g})(r_{\rm g} - r_{4\rm g})}$. We notice that the spin-connection term $\mathcal S^r$ vanishes when the referential geodesic is a spherical orbit, while its
first derivative with respect to $r_{\rm g}$ drastically simplifies and reduces to
\begin{align} 
& 2 \Delta^2 \frac{\partial \mathcal S^r}{\partial r_{\rm g}} =  -\frac{2P_r(r_{\rm g})Y^2_{r\rm g}(r_{\rm g})}{\sqrt{K_{\rm g}} \Sigma}\bigg( \chi_\parallel r_{\rm g} \frac{K_{\rm g} - a^2 z^2_{\rm g}}{K_{\rm g}+ r^2_{\rm g}} \nonumber \\
&-\chi_\perp a z_{\rm g}  \sqrt\frac{K_{\rm g} - a^2 z^2_{\rm g}}{K_{\rm g}+ r^2_{\rm g}} \cos(\psi) \bigg) \,. 
\end{align}

Eq.~\eqref{eq:constraint_spherical_correction} implies that it is not possible to set all corrections to the constants of motion $\delta C_i$ to zero. In other words, it is not possible to define a referential geodesic for a spherical orbit whose constants of motion are the same as those of a spinning trajectory. The second constraint~\eqref{eq:constraint_spherical_correction-derivative} tells us that it is not possible to set to zero the shift to the radial motion at all times, since the shifts $\delta C_i$ are constants.

Thus, spherical orbits, strictly speaking, do not exist for spinning particles. However, with a judicious choice of the constants of motion shifts, the radial correction shift $\delta r$ can be set to zero \textit{on average}, i.e. $\langle \delta r \rangle =0$. We will show in Sec. \ref{ref:fix_the_gauge} how to achieve this.

The general expression for the radial shift, valid for any parametrization, is given by
\begin{align}
\delta r_\parallel &= \frac{1}{2Y^2_{r\rm g}}\bigg(  \displaystyle \sum^{3}_{i=1}\frac{\partial R'_{\rm g}(r_{\rm g})}{\partial C_{i\rm g}} \delta C_i  + \frac{\partial}{\partial r_{\rm g}} R^{\text{sep}}_{\rm s}(r_{\rm g}) \bigg) \nonumber \\
&\phantom{=}  -\frac{P_r(r_{\rm g})r_{\rm g}}{\sqrt{K_{\rm g}} \Sigma} \frac{K_{\rm g} - a^2 z^2_{\rm g}}{K_{\rm g}+ r^2_{\rm g}} \,, \label{eq:rad_shift_parallel_DH_gauge}\\
\delta r_\perp &= \frac{P_r(r_{\rm g})a z_{\rm g}}{\sqrt{K_{\rm g}} \Sigma} \sqrt\frac{K_{\rm g} - a^2 z^2_{\rm g}}{K_{\rm g}+ r^2_{\rm g}} \cos(\psi) \,.
\end{align}

\subsubsection{Polar motion}

The polar motion of nearly spherical orbits has a leading $\mathcal{O}(1)$ geodesic part and a subleading spin correction. It is described by Eq. (46b) of Ref.~\cite{Witzany:2019}:
\begin{equation}
\qty( \frac{\dd z}{\dd \lambda} )^{\! 2} = Z_{\rm g}(z_{\rm g}) + q \delta Z(r_{\rm g},z_{\rm g},\psi)  + \mathcal O(q^2 \chi^2) \,, \label{eq:polar_motion}
\end{equation}
while the equation of motion for the linear spin-correction is
\begin{equation}
   \frac{\dd\delta z}{\dd\lambda} = \pm \frac{\delta Z}{2\sqrt{Z_{\rm g}(z_{\rm g})}} \, , \label{eq:lin-corr-polar-motion} \\
\end{equation}
with
\begin{align}
  \delta Z &= Z_{\rm s} + \frac{\partial Z_{\rm g}}{\partial z_{\rm g}} \delta z + \displaystyle \sum^3_{i =1} \frac{\partial Z_{\rm g}}{\partial C_{i \rm g}} \delta C_{i} \, , \label{eq:diff_polar_potential} \\
  Z_{\rm s} &= Z^{\text{sep}}_{\rm s}(z_{\rm g}) + 2(1 - z^2_{\rm g}) \mathcal S^z \, , \\
  Z^{\text{sep}}_{\rm s}(z_{\rm g}) &= 2(1 - z^2_{\rm g}) \big(\Psi_z(z_{\rm g}) - \delta K_{\rm{eq}} \big)
\end{align}

The spin-connection term $\mathcal S^z = \mathcal S^z(r_{\rm g},z_{\rm g},\psi)$ is given in Appendix C of~\cite{Piovano:2024} with $v^{\rm g}_r =0$ for spherical referential geodesics. The term $\delta K_{\rm{eq}}$ ensures that $\delta z$ is finite in the limit of equatorial orbits. We can write
\begin{align}
z(\wmean) &= z_{\rm g}(w_z)  + q\delta z(\wmean) \, ,
\end{align}
where $z_{\rm g}(w_z)$ is the geodesic polar motion, while
\begin{align}
\delta z(\wmean) &=z_{1{\rm s}\perp}(\wmean) + z_{1{\rm s}\parallel}\sin{\chi_z(w_z)}  - \frac{\dd z_{\rm g}}{\dd w_z}\xi_z(\wmean) \,. \label{eq:spin_polar_trajectory} 
\end{align}
$\wmean = (w_z, w_{\rm p})$ are the mean anomalies that satisfy the equations of motion
\begin{align}
    \frac{\dd w_z}{\dd\lambda} &= \Upsilon_{z\rm g} + q \chi_\parallel \delta \Upsilon_{z}   \, ,  \\
    \frac{\dd w_{\rm p}}{\dd \lambda} &= \Upsilon_{\rm p} \, ,
\end{align}
while $\delta \Upsilon_{z}$ is the correction to the polar frequency. The function $\xi_z(\wmean)$ in Eq. \eqref{eq:spin_polar_trajectory} is given by the infinite sum
\begin{align}
   & \xi_z = \displaystyle \sum_{k = -\infty}^{\infty} \sum_{j = -1}^{1} \frac{i e^{-i \vec \kappa \cdot \wmean}}{\vec \kappa \cdot \vec \Upsilon_{\rm g}}\left(\frac{\delta Y_z}{Y_{z \rm g}}\right)_{\!kj}   \, ,  \label{eq:xi_z} \\
   & \! \left(\frac{\delta Y_z}{Y_{z \rm g}}\right)_{\!kj} = \frac{1}{(2\pi)^2} \int_{(0,2\pi]^2} \dd^2 \mathsf{w} \frac{\delta Y_y(\wmean)}{Y_{y \rm g}(w_y)} e^{i \vec \kappa \cdot \wmean} \,,
\end{align}
where $\vec\Upsilon_{\rm g} = (\Upsilon_{z\rm g}, \Upsilon_{\rm p})$ and $\vec \kappa = (k, j)$ with $j \in (-1,0,1)$. $\xi_z$ is not defined when the geodesic trajectory \textbf{satisfies the resonance condition $\vec \kappa \cdot \vec \Upsilon_{\rm g} = 0$. However, such a condition is never met at the linear-in-spin order, since $j=\pm 1$ and $0.67 \lesssim \Upsilon_{\rm p}/\Upsilon_{z{\rm g}} < 1$ across the parameter space}\footnote{This was numerically checked across the parameter space $a \in [0, 1]M$, $p \in [r_\text{ISSO}, +\infty)$, $x \in [-1, 1]$.}. \textbf{Thus, Eq.~\eqref{eq:xi_z} is always valid.} 
The correction to the polar frequency is given by the average
\begin{equation}
  \delta \Upsilon_{z} = \Upsilon_{z\rm g}\left \langle\frac{\delta Y_z}{Y_{z \rm g}} \right\rangle \equiv \Upsilon_{z\rm g}\left(\frac{\delta Y_z}{Y_{z \rm g}}\right)_{\!00}   \,.
\end{equation}
which can be written in analytic form in terms of Legendre elliptic integrals (see Supplemental Material \cite{SupMat}).

Finally, the shift to the turning points $z_{1{\rm s}\parallel}$ is constant and is proportional to $\chi_\parallel$, while $z_{1{\rm s}\perp}(\psi_{\rm p})$ is proportional to $\chi_\perp$ and depends on the spin precession angle $\psi$ (see Appendix D of~\cite{Piovano:2024}).

\subsubsection{Coordinate-time and azimuthal motion}

Once we solve the equations of motion for the radial and polar trajectories, the coordinate-time and azimuthal trajectories can be efficiently computed in terms of Fourier series in the mean anomalies. We first expand the geodesics coordinate-time and azimuthal velocities $\dd t_{\rm g}/\dd\lambda$ and $\dd\phi_{\rm g}/\dd\lambda$, respectively,
\begin{align}
\frac{\dd t_{\rm g}}{\dd\lambda} &= V^t_{r\rm g} + \displaystyle \sum_{k = -\infty}^{\infty}(V^t_{z\rm g})_k e^{-i k w_z}  \, , \\
\frac{\dd \phi_{\rm g}}{\dd\lambda} &= V^\phi_{r\rm g} + \displaystyle \sum_{k = -\infty}^{\infty}(V^\phi_{z\rm g})_k e^{-i k w_z} \, ,
\end{align}
and the corresponding spin corrections to the velocities
\begin{align}
\frac{\dd \delta t}{\dd\lambda} &= \displaystyle \sum_{\vec \kappa}(\delta V^t)_{kj} e^{-i\vec \kappa \cdot \wmean}  \, , \label{eq:dtsdlFour} \\
\frac{\dd \delta\phi}{\dd\lambda} &= \displaystyle \sum_{\vec \kappa}(\delta V^\phi)_{kj} e^{-i\vec \kappa \cdot \wmean} \, , \label{eq:dphisdlFour}
\end{align}
where we introduced the notation
\begin{equation}
\displaystyle \sum_{\vec \kappa} \equiv \displaystyle \sum_{k = -\infty}^{\infty} \sum_{j = -1}^{1} \,.
\end{equation}
The geodesics terms $V^t_{r\rm g}$, $V^\phi_{r\rm g}$, $V^t_{z\rm g}$, and $V^\phi_{z\rm g}$ are given in Sec.~\ref{ref:geodesic_motion}, while the variations $\delta V^t$ and $\delta V^\phi$ are given by
\begin{align}
  \delta V^t &= -\frac{\Sigma}{2}\Gamma^t + \frac{\partial V^t_{r \rm g}}{\partial r_{\rm g}} \delta r + \frac{\partial V^t_{z\rm g}}{\partial z_{\rm g}} \delta z +\displaystyle \sum^2_{i =1} \frac{\partial V^t_{\rm g}}{\partial C_{i \rm g}} \delta C_i \, ,  \label{eq:diff_time_potential} \\
  \delta V^\phi &= -\frac{\Sigma}{2}\Gamma^\phi + \frac{\partial V^\phi_{r \rm g}}{\partial r_{\rm g}} \delta r +  \frac{\partial V^\phi_{z \rm g}}{\partial z_{\rm g}} \delta z +\displaystyle \sum^2_{i =1} \frac{\partial V^\phi_{\rm g}}{\partial C_{i \rm g}} \delta C_i  \, , \label{eq:diff_azimuthal_potential}
\end{align}
where $V^t_{\rm g} = V^t_{r\rm g} + V^t_{z\rm g}$, $V^\phi_{\rm g} = V^\phi_{r\rm g} + V^\phi_{z\rm g}$. 
The terms $\Gamma^t$ and $\Gamma^\phi$ are the projections of the Marck tetrad into the Christoffel symbol $\Gamma^\mu_{\rho \sigma}$, and can be found in Appendix C of~\cite{Piovano:2024}. After integrating the velocities
\begin{align}
\frac{\dd t}{\dd\lambda} &= \frac{\dd t_{\rm g}}{\dd\lambda} + q \frac{\dd\delta t}{\dd\lambda}  \, , \\
\frac{\dd\phi}{\dd\lambda} &= \frac{\dd\phi_{\rm g}}{\dd\lambda} + q \frac{\dd\delta \phi}{\dd\lambda} 
\end{align}
we get
\begin{align}
t(\lambda) &= \Big(\Upsilon_{t \rm g} + q \chi_\parallel \delta\Upsilon_t \Big)\lambda + \Delta t_{\rm g}(\lambda) + q \chi \delta\Delta t(\lambda) \, \label{eq:Fourier_series_solution_t} \ ,\\
\phi(\lambda) &= \Big(\Upsilon_{\phi\rm g} + q \chi_\parallel \delta\Upsilon_\phi \Big)\lambda + \Delta \phi_{\rm g}(\lambda)  + q \chi \delta\Delta \phi(\lambda) \label{eq:Fourier_series_solution_phi}
\end{align} 
with $\delta \Upsilon_t$ and $\delta \Upsilon_\phi$ the spin corrections to the Mino-time frequencies. 
Analytic expressions in terms of Legendre elliptic integrals are presented in Supplemental Material \cite{SupMat}. The terms $\Delta t_{\rm g}(\lambda)$ and $\Delta\phi_{\rm g}(\lambda)$ are the purely oscillatory components of the geodesic solutions, 
\begin{align}
\Delta t_{\rm g}(\lambda) &= \sum_{k \neq 0} \frac{i(V^t_{z\rm g})_k}{k (\Upsilon_{z\rm g})} e^{-i k w_z}  \, , \label{eq:geotFourierseriesosc} \\
\Delta \phi_{\rm g}(\lambda) &= \sum_{k \neq 0} \frac{i(V^\phi_{z \rm g})_k}{k (\Upsilon_{z\rm g})} e^{-i k w_z} \ ,  \label{eq:geophiFourierseriesosc}
\end{align} 
whereas $\Delta \delta t(\lambda)$ and $\Delta \delta \phi(\lambda)$ are the corresponding spin corrections given by
\begin{align}
\delta\Delta t(\lambda) &= \frac{\delta \Upsilon_{z}}{\Upsilon_{z \rm g}}\Delta t_{\rm g}(\lambda) + \displaystyle \sum_{\vec \kappa \neq 0} \frac{i(\delta V^t)_{kj}}{ \vec \kappa \cdot \vec \Upsilon_{\rm g}} e^{-i \vec \kappa \cdot \wmean} \, , \label{eq:spintFourierseriesosc} \\
\delta\Delta\phi(\lambda) &= \frac{\delta \Upsilon_{z}}{\Upsilon_{z \rm g}}\Delta \phi_{\rm g}(\lambda) + \displaystyle \sum_{\vec \kappa \neq 0} \frac{i(\delta V^\phi)_{kj}}{\vec \kappa \cdot \vec \Upsilon_{\rm g}} e^{-i \vec \kappa \cdot \wmean}  \, . \label{eq:spinphiFourierseriesosc}
\end{align} 

\subsubsection{Fixing the spin gauge} \label{ref:fix_the_gauge}

The linearized MPD equations admit a phase-space gauge freedom, which is encoded in the freedom of choosing a referential geodesic \cite{Witzany:2019,Piovano:2024}. It was demonstrated in Ref.~\cite{Mathews:2025} that the choice of the referential geodesic used to parametrize the motion of a spinning test body does not affect other first-order self-force effects, at least at order $\mathcal O (\chi q)$. Henceforth, we label ``spin gauge'' the specific referential geodesic adopted for the calculation of the secondary spin trajectories and the corresponding shifts to the constants of motion and frequencies. In the Hamilton-Jacobi formalism, selecting a spin gauge is equivalent to determining the appropriate shifts to the constants of motion $\delta C_i$.%
\footnote{In some cases, it may be necessary to fix other orbital elements. For instance, in order to impose that a fiducial geodesic matches the initial condition of a spin-perturbed orbit, one needs to find the appropriate initial phase shift of the mean anomalies.} %
Once the shifts $\delta C_i$ and $\delta r$ have been determined, the equations of the previous sections are uniquely fixed.

There are a number of choices to fix the spin gauge, for an overview see Ref. \cite{Piovano:2024}. Here we use the ``fixed turning point'' parametrization introduced in Refs.~\cite{Drummond:2022xej,Drummond:2022efc}. This means that we require that the average turning points of the spinning particle trajectory are the same as those of the referential geodesic. We choose the referential geodesic to be a spherical orbit, so the spinning particle trajectory is also a spherical orbit in an average sense. In other words, we have that $\langle \delta r \rangle = 0$, where $\langle \rangle$ denotes an average over orbital phases. 

We determine the corresponding spin corrections $\delta E, \delta L_z, \delta K$ by imposing that the averages of the spin corrections to the radial potential~\eqref{eq:diff_radial_potential} and polar potential~\eqref{eq:diff_polar_potential} are zero at the geodesics' turning points $r_{\rm g}$ and $z_{1\rm g}$, which leads to the conditions
\begin{subequations}\label{eq:condition_turning_points_linear}
\begin{align} 
    \langle \delta R(r_{\rm g},z_{\rm g},\psi) \rangle &= 0 \ ,\\
    \langle \partial_{r_{\rm g}} \delta R(r_{\rm g},z_{\rm g},\psi) \rangle &= 0 \ , \\
    \langle \delta Z(r_{\rm g},z_{1 \rm g},\psi) \rangle &= 0 \ ,
\end{align}
\end{subequations}
which implies  
\begin{subequations}\label{eq:system_shift_const_motion_DH_gauge}
\begin{align} 
&R^{\text{sep}}_{\rm s}(r_{\rm g})  + \displaystyle \sum^{3}_{i=1}\frac{\partial R_{\rm g}}{\partial C_{i\rm g}} \delta C_i =0 \ , \\
&\frac{\partial R^{\text{sep}}_{\rm s}}{\partial r_{\rm g}} +  2 \Delta^2 \left \langle\frac{\partial \mathcal S^r}{\partial r_{\rm g}} \right \rangle +  \displaystyle \sum^{3}_{i=1}\frac{\partial R'_{\rm g}(r_{\rm g})}{\partial C_{i\rm g}} \delta C_i =0 \ , \\
& Z^{\text{sep}}_{\rm s}(z_{1\rm g}) + 2(1 - z^2_{1\rm g}) \langle \mathcal S^z \rangle + \displaystyle \sum^3_{i =1} \frac{\partial Z_{\rm g}}{\partial C_{i \rm g}} \delta C_i = 0 \, ,
\end{align}
\end{subequations}
with the averages $ \langle\partial \mathcal S^r /\partial r_{\rm g} \rangle $ and $ \langle \mathcal S^z \rangle $ given by
\begin{align}
  \left \langle\frac{\partial \mathcal S^r}{\partial r_{\rm g}} \right \rangle  & = \frac{r_{\rm g}P_r(r_{\rm g})Y^2_{r\rm g}(r_{\rm g})}{\sqrt{K_{\rm g}}(K_{\rm g} + r^2_{\rm g})\Delta^2}  \bigg(1 - \frac{K_{\rm g} + r^2_{\rm g}}{r^2_{\rm g} \mathsf{K}(k_z)} \mathsf{\Pi}(\gamma_r| k_z) \bigg) \ , \\
  \langle \mathcal S^z \rangle & = - \frac{(K_{\rm g} + r^2_{\rm g})z^2_{1\rm g}Y^2_{z\rm g}(z_{1\rm g})P_{z}(z_{1\rm g})}{\sqrt{K_{\rm g}}(1-z^2_{1 \rm g})(K_{\rm g} - a^2 z^2_{1 \rm g})(r^2 + a^2 z^2_{1 \rm g})} \, \ .
\end{align}
Here we have defined
\begin{align}\label{eq:kz_gammar}
k_z & \equiv a^2 (1 - E_{\rm g}^2) \frac{z^2_{1\rm g}}{z^2_{2\rm g}} \ ,  \qquad \gamma_r \equiv - a^2\frac{z^2_{1\rm g}}{r^2_{\rm g}} \ ,
\end{align}
while $\mathsf{K}(k_z)$ and $\mathsf{\Pi}(\gamma_r| k_z)$ are the complete elliptic integrals of the first and third kind, respectively. We follow the \textit{Wolfram Mathematica} convention for the elliptic integrals.\footnote{Note that in the conventions of \textit{Mathematica} the argument $k_z$ is \textit{not} the elliptic modulus but rather its square, which would often be denoted as $m_z$ in the mathematics literature. However, here we follow the notation of \citet{vandeMeent:2020}.}

We now define the following matrix
\begin{equation}
    \mathbb{M} \equiv \mqty( \pdv{R_\text{g}}{E_{\rm g}} & \pdv{R_\text{g}}{L_{z \rm g}} & \pdv{R_\text{g}}{K_{\rm g}} \\ \pdv{R'_\text{g}}{E_{\rm g}} & \pdv{R'_\text{g}}{L_{z \rm g}} & \pdv{R'_\text{g}}{K_{\rm g}} \\ \pdv{Z_{\rm g}}{E_{\rm g}} & \pdv{Z_{\rm g}}{L_{z \rm g}} & \pdv{Z_{\rm g}}{K_{\rm g}} )
\end{equation}
and vector 
\begin{align}
  \vec{\mathrm v}_{\rm s} &= \mqty( R^{\text{sep}}_{\rm s}(r_{\rm g}) \\ \frac{\partial R^{\text{sep}}_{\rm s}}{\partial r_{\rm g}} +  2 \Delta^2 \left \langle\frac{\partial \mathcal S^r}{\partial r_{\rm g}} \right \rangle \\ Z^{\text{sep}}_{\rm s}(z_{1\rm g}) + 2(1 - z^2_{1\rm g}) \langle \mathcal S^z \rangle ) \,,
\end{align}
which allow us to solve the linear system~\eqref{eq:system_shift_const_motion_DH_gauge} with unknowns $( \delta E , \delta L_z , \delta K )$  as
\begin{align}
    \mqty( \delta E \\ \delta L_z \\ \delta K ) &= - \mathbb{M}^{-1}\vec{\mathrm v}_{\rm s} \,. \label{eq:dC}
\end{align}

Alternatively, the shifts can be obtained by carefully expanding in the eccentricity $e$ the expression given in Eq.~(3.85) of Ref.~\cite{Piovano:2024}. Indeed, despite the radial shifts to the turning points diverging for $e\to 0$ in the Hamilton-Jacobi formalism, the spin corrections for the constants of motion  are finite for $e\to 0$. We numerically checked that both methods lead to the same results. Their implementation, together with the analytic corrections to the Mino-time frequencies, is available in Supplemental Material \cite{SupMat}.

\begin{figure}[t!]
    \centering
    \includegraphics[width=0.75\linewidth]{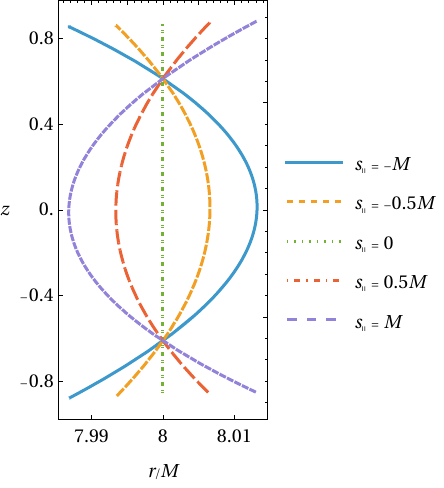}
    \caption{Projection of the near-spherical trajectories into the $r-z$ plane for different values of $s_\parallel$ with fixed turning points. The orbital parameters are $a = 0.95 M$, $p=8M$, and $x=0.5$. The secondary spin causes a slight bending of the orbit away from a trajectory of constant radius, which led us to call this class of trajectories only ``nearly spherical''.  (The value of $s_\parallel$ is unphysically large to make the cases distinguishable; typical EMRI parameters would show much smaller effects.)}
    \label{fig:trajectory_r_z_plane}
\end{figure}

In Fig. \ref{fig:trajectory_r_z_plane} we show the projection of the trajectory into the $r-z$ plane for $s_\perp = 0$ and multiple values of $s_\parallel$. The trajectory ``bends'' very slightly toward or away from the black hole for different inclinations. This could possibly be interpreted as the spins becoming slightly inclined with respect to each other, thus causing a shift in the spin-spin term in an effective-potential interaction.

\subsubsection{The innermost stable spherical orbit}

In the fixed-turning-point gauge, it is also possible to derive the location of the innermost stable spherical orbit (ISSO), i.e., the linear-in-spin correction to the value of $p$ that separates bound and plunging orbits. 

\begin{figure}[t!]
    \centering
    \includegraphics[width=\linewidth]{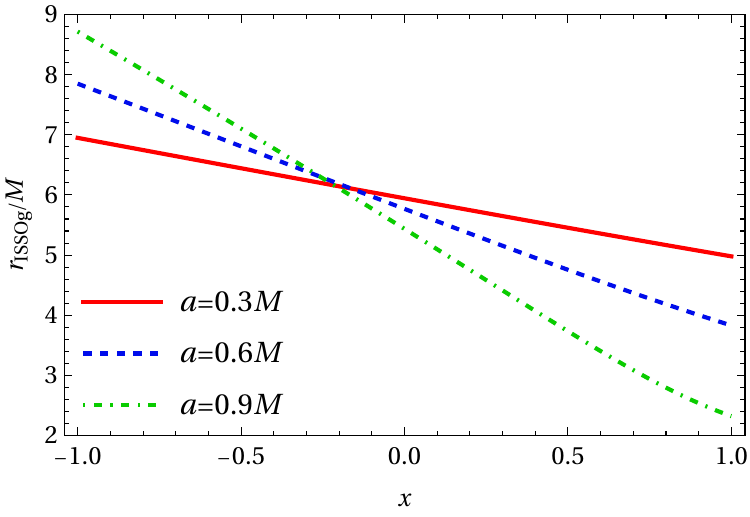}
    \includegraphics[width=\linewidth]{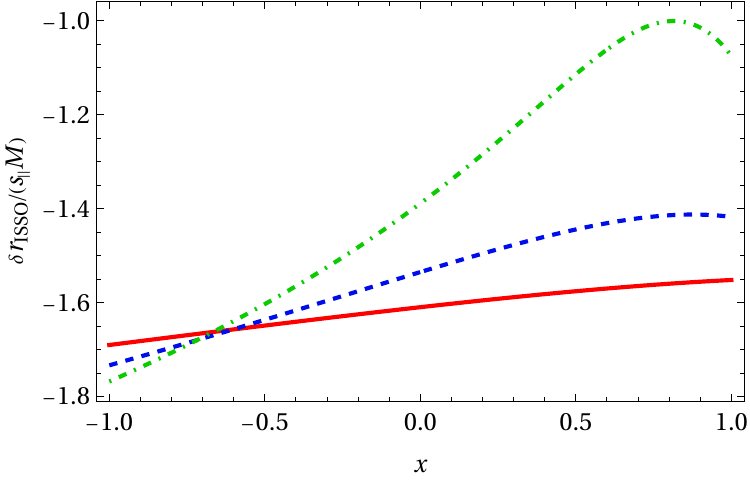}
    \caption{Radius of the innermost spherical geodesics $r_{\rm ISSOg}$ and the secondary spin correction $\delta r_{\rm ISSO}$ as a function of the inclination parameter $x$ for several values of primary black hole spin.}
    \label{fig:rISSO}
\end{figure}

The derivation relies on the analytical solution of the equations of motion introduced in Ref. \cite{Skoupy:2024uan} and the analysis of the innermost spherical geodesics by \citet{Stein:2019buj}; we present it in full in Appendix \ref{app:rISSO}. The resulting formula reads 
\begin{align}
     r_\text{ISSO}(x) = & r_{\rm ISSO,g}(x_{\rm g}) + \delta r_\text{ISSO}(x_{\rm g})\,, \label{eq:rISSO}\\
    \begin{split}
    \delta r_\text{ISSO}(x_{\rm g}) = 
    & -\frac{\partial r_{\rm ISSO,g}}{\partial x_{\rm g}} s_\parallel \frac{a(1 - x^2_{\rm g}) \left(L_{z\rm g} - a E_{\rm g} x^2_{\rm g} \right)}{x_{\rm g} \sqrt{K_{\rm g}} \left(r^2_{\rm ISSO,g} + a^2 (1-x^2_{\rm g})\right)} 
    \\ & - s_\parallel \frac{(r^2_{\rm ISSO,g} + a ^2) E_{\rm g} - a L_{z\rm g}}{r_{\rm ISSO,g}\sqrt{K_{\rm g}}} \frac{\mathsf{\Pi}(\gamma_r|k_z)}{\mathsf{K}(k_z)} \,.
    \end{split} \label{eq:rISSOshift}
\end{align}
The $\partial r_{\rm ISSO,g}/\partial x_{\rm g}$ was then computed from the implicit function theorem and the separatrix polynomial of \citet{Stein:2019buj}. Plots of the geodesic ISSO $r_\text{ISSO}(x_{\rm g})$ and $\delta r_\text{ISSO}(x_{\rm g})$ are given in Fig.~\ref{fig:rISSO}. We numerically checked that $\delta r_\text{ISSO}$ converges to the correct value $-s_\parallel \sqrt{8/3}$ for Schwarzschild space-time and $0$ for extremal Kerr black holes in the $x\to 1$ circular limit \cite{Jefremov:2015gza}.

\section{GW fluxes}\label{sec:GW_fluxes}

GW emission drives the orbital evolution \textbf{and} makes EMRIs observable. We calculate the GW fluxes using the Teukolsky equation, which captures the full relativistic strong-field dynamics of GWs in Kerr spacetime. Specifically, we compute fluxes sourced by nearly spherical orbits of spinning particles presented in the previous Sections, and we employ the Teukolsky equation in the frequency domain. The calculation is based on the formalism derived in \cite{Skoupy:2023} for fluxes from generic orbits with an additional step in which the fluxes are split into the geodesic part and the linear-in-spin part similarly to Ref. \cite{Piovano:2024}.

\subsection{Flux computation}\label{sec:flux_computation}

We start by expressing the fluxes of specific energy and angular momentum normalized by the mass ratio $\mathcal{F}^E$ and $\mathcal{F}^{L_z}$ in terms of Teukolsky amplitudes per unit mass $C^\pm_{lmk}$ as
\begin{align}
    \mathcal{F}^E &= \sum_{lmk} \mathcal{F}^E_{lmk} \, , \quad \mathcal{F}^{L_z} = \sum_{lmk} \mathcal{F}^{L_z}_{lmk}  \,, \label{eq:fluxes}\\
    \mathcal{F}^E_{lmk} &= M \frac{\abs{C^+_{lmk}}^2 + \alpha_{lmk} \abs{C^-_{lmk}}^2}{4 \pi \omega_{mk}^2} \,, \label{eq:fluxes_E}\\
    \mathcal{F}^{L_z}_{lmk} &= m M \frac{\abs{C^+_{lmk}}^2 + \alpha_{lmk} \abs{C^-_{lmk}}^2}{4 \pi \omega_{mk}^3} \,, \label{eq:fluxes_L_z}
\end{align}
where $l$, $m$, and $k$ are the mode numbers, the frequency $\omega_{mk}$ reads
\begin{equation}
    \omega_{mk} = m \Omega_\phi + k \Omega_z
\end{equation}
and $\alpha_{lmk}$ is a constant given in Eq.~(57) of Ref.~\cite{Skoupy:2023}. 

An important property of the linearized motion is that the fluxes are independent of the orthogonal component $\chi_\perp$. This stems from the fact that only the $j = \pm 1$ amplitudes depend on $\chi_\perp$ and, since they behave as $\order{s}$, they do not have to be included in the fluxes \textbf{(see Sec. \ref{sec:flux_derivatives})}. %
Our calculations therefore hold for any configuration of the secondary spin.

The amplitudes can be also used to calculate the snapshot waveform
\begin{equation}
    h = h_+ - i h_\times = -\frac{2\mu}{r} \sum_{lmk} \frac{C^+_{lmk}}{\omega_{mk}^2} {}_{-2}S_{lm}^{a\omega_{mk}}(\theta) e^{-i \omega_{mk} t + i m \phi} \, ,
\end{equation}
where ${}_{-2}S_{lm}^{a\omega_{mk}}(\theta)$ is the spin-weighted spheroidal harmonic. \textbf{The waveform also has a contribution from the $j=\pm 1$ amplitudes with frequencies $\omega_{mkj} = \omega_{mk} + j \Omega_{\rm p}$, which arise from the precession of secondary spin. However, these amplitudes are $\order{s_\perp}$,  and are almost certainly unobservable in the case of EMRIs~\cite{Burke:2023lno}. However, they may be relevant in the case of intermediate mass-ratio inspirals.} 

The amplitudes and spin-weighted spheroidal harmonics satisfy the symmetry conditions
\begin{subequations}\label{eq:symmetries_C_S}
\begin{align}
    C^\pm_{l,-m,-k} &= (-1)^{l+k} \overline{C}^\pm_{lmk} \,, \\
    {}_{-2}S_{l,-m}^{-a\omega}(\theta) &= (-1)^{l} {}_{-2}S_{lm}^{a\omega}(\pi-\theta) \,,
\end{align}
\end{subequations}
which will be useful in the numeric calculation in Sec. \ref{sec:FD_waveforms}.

The asymptotic Teukolsky amplitudes are calculated as
\begin{equation}\label{eq:Cpm_lmk}
    C^{\pm}_{lmk} = \frac{1}{W_{lmk} \Upsilon_t} \int_0^{2\pi} \dd w_z \; I^{\pm}_{lmk}(w_z) e^{i \varphi_{mk}(w_z) + i k w_z} \, ,
\end{equation}
where $W_{lmk}$ is the invariant Wronskian defined below Eq. (46) in \cite{Skoupy:2023} and
\begin{equation}
    \varphi_{mk}(w_z) = \omega_{mk} \Delta t(w_z) - m \Delta \phi(w_z) \, .
\end{equation}
The function $I^\pm_{lmk}$, which contains the dependence on the trajectory and the solutions of homogeneous radial and angular Teukolsky equation, can be split into contributions from the particle's monopole and dipole as
\begin{align}
    I^\pm_{lmk} &= I^{\text{m}\pm}_{lmk} + I^{\text{d}\pm}_{lmk} \,, \\
    I^{\text{m}\pm}_{lmk} &= \Sigma \sum_{ab} \qty( u_a u_b F^{ab\pm}_{lmk} ) \,, \\
    I^{\text{d}\pm}_{lmk} &= \Sigma \sum_{ab} \qty( B^0_{ab} F^{ab\pm}_{lmk} + B^z_{ab} \pdv{F^{ab\pm}_{lmk}}{z} + B^r_{ab} \pdv{F^{ab\pm}_{lmk}}{r} ) \,,
\end{align}
where $ab = nn, n\Bar{m}, \Bar{m} \Bar{m}$,
\begin{align}
    F^{ab\pm}_{lmk} &= \sum_{i=0}^{I_{ab}} (-1)^i f^{(i)}_{ab} \dv[i]{R^\mp_{lmk}}{r} \,, \\
    B^0_{ab} &=  A^{\rm d}_{ab} + i \qty( \omega B^t_{ab} - m B^\phi_{ab} ) \,.
\end{align}
The quantities $A^\text{d}_{ab}$, $B^\mu_{ab}$, and $f^{(i)}_{ab}$ are defined in Eqs. (49) and (B4) in \cite{Skoupy:2023}. $R^\mp_{lmk}(r)$ are the homogeneous solutions of the radial Teukolsky equation satisfying ``in'' and ``up'' boundary conditions, respectively \cite{Pound:2021}.

Unlike the fluxes, the amplitudes $C^\pm_{lmk}$ depend on the initial phase of the polar and azimuthal motion. For generic initial values of $z$ and $\phi$, an additional phase factor $\xi_{mk}$ must be included (see Eqs.~(3.18)--(3.20) in \cite{Hughes:2021}). However, the secondary spin contribution to $\xi_{mk}$ does not need to be taken into account since it is a 2PA effect. Therefore, for simplicity, we calculate the amplitudes and waveforms for trajectories starting at the turning point $z(t=0) = \sqrt{1-x^2}$ and leave the fully generic case for future work.

\subsection{Derivatives of the fluxes}\label{sec:flux_derivatives}

To calculate the inspirals, it is convenient to linearize the fluxes as
\begin{equation}
    \mathcal{F}^{E,L_z} = \mathcal{F}^{E,L_z}_\text{g} + q \chi_\parallel \delta \mathcal{F}^{E,L_z}
\end{equation}
and calculate the derivatives of the geodesic fluxes with respect to the orbital parameters $p$ and $x$.
\textbf{The reason why the flux depends only on $\chi_\parallel$ is as follows. The $j = \pm 1$ modes behave as $\order{s_\perp^2} = \order{s^2}$ since the  corresponding amplitudes are proportional to $s_\perp$ and they are squared in Eqs.~\eqref{eq:fluxes_E} and \eqref{eq:fluxes_L_z}.\footnote{This is also apparent in Eqs. (4.53), (4.74), and (4.75) of \cite{Piovano:2024}. The leading order partial amplitudes only include contributions from the geodesic orbits, therefore they vanish for $j=\pm 1$} On the other hand, the orthogonal part of the secondary spin only causes an oscillating contribution, which does not contribute to $j = 0$ Fourier mode (this is a general property of fluxes from all bound orbits, as discussed in Refs. \cite{Skoupy:2023,Piovano:2024}).
Consequently, we can expand the $j = 0$ amplitudes in spin as} 
\begin{equation}
    C^{\pm}_{lmk} = C^{\pm \text{g}}_{lmk} + q \chi_\parallel \delta C^{\pm}_{lmk} \,,
\end{equation}
where we calculate the linear part $\delta C^{\pm}_{lmk}$ by linearizing the formula \eqref{eq:Cpm_lmk}. Then, the linear part can be calculated as
\begin{multline}\label{eq:deltaCpm_lmk}
    \delta C^\pm_{lmk} = -\qty( \frac{\delta \Upsilon_t}{\Upsilon_{t\text{g}}} + \frac{\delta W_{lmk}}{W_{lmk}^\text{g}}) C^{\pm\text{g}}_{lmk} \\ 
    + \frac{1}{W_{lmk}^\text{g} \Upsilon_{t\text{g}}} \int_0^{2\pi} \qty(\delta I^\pm_{lmk} + i I^{\pm\text{g}}_{lmk} \delta \varphi_{mk} ) e^{i \varphi_{mk}^\text{g}(w_z)} \dd w_z \, ,
\end{multline}
where
\begin{align}
    \delta \psi_{mk} &= \delta\omega_{mk}^{\text{g}} \Delta t_\text{g} + \omega_{mk} \delta\Delta t - m \delta\Delta\phi \,.
\end{align}
Since the dipole contribution $I^{\text{d}\pm}_{lmk}$ is proportional to spin, we can simply calculate it with geodesic quantities. The monopole contribution can be linearized as
\begin{align}
    \delta I^{\text{m}\pm}_{lmk} &= \Sigma_{\text{g}} \bigg( u_{a\text{g}} u_{b\text{g}} \qty( \frac{\delta \Sigma}{\Sigma_{\text{g}}} F^{ab\pm\text{g}}_{lmk} + \delta F^{ab\pm}_{lmk} ) \nonumber\\
    &\phantom{=} + 2 u_{(a\text{g}} \delta u_{b)} F^{ab\pm\text{g}}_{lmk} \bigg) \,,
\end{align}
where
\begin{align}
    \delta \Sigma &= 2 ( r_\text{g}  \delta r + a^2 z_\text{g} \delta z ) \,, \\
    \delta u_b &= \delta u_\mu \lambda^\mu_b + u_{\mu\text{g}} \partial_y \lambda^\mu_b \delta y \,, \\
    \delta F^{ab\pm}_{lmk} &= \partial_y F^{ab\pm\text{g}}_{lmk} \delta y + \partial_\omega F^{ab\pm\text{g}}_{lmk} \delta \omega_{mk} \,,
\end{align}
with $y=r,z$ and $\lambda^\mu_b$ is the Kinnersley tetrad. To calculate $\partial_\omega F^{ab\pm\text{g}}_{lmk}$, we need to find the derivatives of the radial functions $R^\pm_{lm\omega}(r)$ and spin-weighted spheroidal harmonics ${}_{-2}S_{lm}^{a\omega}$ with respect to the frequency. For this, we use the method developed in \cite{Piovano:2021}, where the solution by \citet{Leaver:1985ax} is expanded in spin and the expanded radial Teukolsky equation is numerically solved with hyperboloidal slicing. 

The derivatives of the geodesic amplitudes with respect to $p$ and $x$ can be calculated similarly by differentiating formula \eqref{eq:Cpm_lmk}. The result can be expressed by replacing $\delta$ with $\partial_{r,x}$ in the previous equations. The derivatives of the trajectory can be found by differentiating the analytical formulas from \cite{vandeMeent:2020} \textbf{and using identities for derivatives of the elliptic integrals and Jacobi elliptic functions \cite{NIST:DLMF}}.

\section{Flux-driven inspirals}\label{sec:inspirals}
In this Section we establish the evolution of the orbital elements from the GW fluxes computed in the previous Sections. First we prove that near-spherical orbits stay near-spherical also for the radiation reaction on spinning test particles, and then we describe how we use this to evolve the orbital elements and produce the corresponding waveforms.

\subsection{Do spherical inspirals stay spherical?}

Intuitively, eccentric orbits can be viewed as ``more excited'' than circular or spherical ones. A circular or spherical orbit can then be viewed as a ``quasi-equilibrium state''. As such, we expect that circular and spherical orbits do not spontaneously develop eccentricity under radiation-reaction. Is this indeed the case? Additionally, the notion of eccentricity is coordinate dependent, and one needs to find in which coordinates the zero-eccentricity orbits are fixed under radiation-reaction. 
 
For geodesics, \citet{Kennefick:1995za} showed that spherical orbits in the sense of constant Boyer-Lindquist radius in Kerr space-time stay spherical under adiabatic radiation-reaction. The restriction to constant radius means that one of the three constants of motion $E_{\rm g},L_{z \rm g},K_{\rm g}$ becomes dependent. As a result, one is able to evolve spherical inspirals of non-spinning particles just from the condition of sphericity and the energy and angular momentum fluxes $\mathcal{F}^E, \mathcal{F}^{L_z}$ through the flux-balance laws
\begin{equation}
    \dv{t} \mqty( E \\ L_z ) = - q \mqty( \mathcal{F}^E \\ \mathcal{F}^{L_z} )
\end{equation}
without the need to compute the evolution of the Carter constant explicitly \cite{Hughes:1999bq,Hughes:2001jr}. 

However, in the case of near-spherical orbits of spinning particles, the argument of Kennefick and Ori applies only to the leading-order $\mathcal{O}(q)$ terms driving the inspiral, and one has to consider the possibility of a growing eccentricity due to the spin contribution to radiation-reaction. From the knowledge of $\mathcal{F}^E, \mathcal{F}^{L_z}$ we know how to evolve energy and angular momentum, which cannot be used to evolve all of the three orbital parameters $p,e,x$. Evolving these would require also evaluating $\langle \dot{K}_{\rm R}\rangle$, which could in principle be done by combining the results of Refs \cite{Isoyama:2018sib,Grant:2024ivt,Witzany:2024ttz} (see also Ref. \cite{Mathews:2025}). We present here a short argument why this is not necessary and why we are able to still evolve the orbit under the near-sphericity constraint $e=\dot{e}=0$. An alternative Kennefick-Ori type argument using the virtual-trajectory formalism of Ref. \cite{Skoupy:2024uan} is also presented in Appendix \ref{app:vtInsp}. 

\begin{figure}
    \centering
    \includegraphics[width=\linewidth]{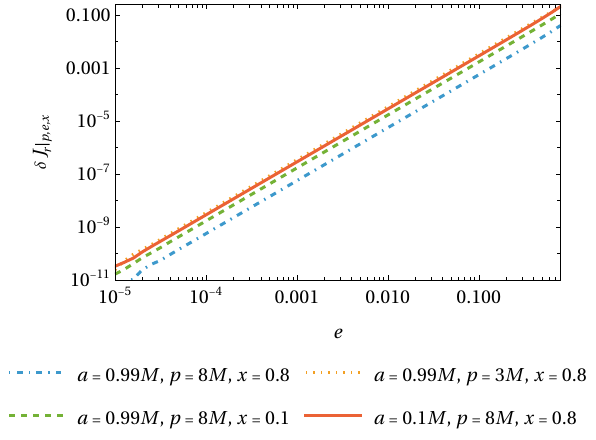}
    \caption{Dependence of the linear part of the radial action with fixed turning points $\eval{\delta J_r}_{p,e,x}$ on eccentricity $e$ for different orbital parameters. }
    \label{fig:deltaJr}
\end{figure}

According to Refs \cite{Isoyama:2018sib,Grant:2024ivt} the flux of the radial action is
\begin{equation}
    \dv{J_r}{t} = - \sum_{lmnk} \frac{n \qty(\abs{C^+_{lmnk}}^2 + \alpha_{lmnk} \abs{C^-_{lmnk}}^2)}{4 \pi \omega_{mnk}^3}\,,
\end{equation}
where $n$ is the harmonic number of the radial motion. This behaves as $\order{e^2}$ (for small $e$, the leading term comes from $n=\pm 1$, where $C^\pm_{lmnk} = \order{e^\abs{n}}$, \textbf{which is derived from the properties of Fourier expansion of the trajectory \cite{Fujita:2009bp} and its subsequent use in Bessel type formulas in the amplitude integrals \cite{peters1964gravitational,Drasco:2006,Sago:2015}}). This holds also for a spinning secondary in the fixed turning points framework \textbf{since the integral in Eq.~(54) of \cite{Skoupy:2023} and the trajectory from \cite{Drummond:2022xej,Drummond:2022efc} have the same structure regarding exponential terms and trajectory dependence}. If we can show that $J_r = \order{e^2}$, then $\dot{e} = \order{e}$ and near-spherical orbits stay near-spherical. The desired behavior of the geodesic part $J_{r\text{g}}$ can be simply verified by changing the integration variable in Eq.~(20) of \cite{Schmidt:2002} to $\chi_r$ from Eq.~\eqref{eq:r_param}. On the other hand, the linear-in-spin part of the radial action with fixed turning points can be calculated from the linear part of the radial action with fixed constants of motion as
\begin{equation}
    \eval{\delta J_r}_{p,e,x} = \eval{\delta J_r}_{C_i} + \pdv{J_{r\text{g}}}{C_{i\text{g}}} \eval{\delta C_i}_{p,e,x}\,,
\end{equation}
where $\eval{\delta J_r}_{C_i}$ \textbf{can be found in Eq.~(35)} of \cite{Witzany:2024ttz} and $\delta C_i = \delta E,\delta L_z,\delta K$ are given in \textbf{Eq.~(3.80)} of \cite{Piovano:2024}. We evaluated this formula at several points of the parameter space and found that $\eval{\delta J_r}_{C_i}$ has a finite value for spherical orbits and numerically checked that $\eval{\delta J_r}_{p,e,x} = \order{e^2}$. \textbf{The result is shown in Fig. \ref{fig:deltaJr} where we can see the quadratic behavior of $\eval{\delta J_r}_{p,e,x}$ in the log-log plot.}

\subsection{Evolution of orbital elements and waveform}

Using the results of the previous Section, we determine the flux-driven evolutions of the orbital parameters $p,x$ as
\begin{equation}\label{eq:rdot_xdot}
    \dv{t} \mqty( p \\ x ) = - q \mathbb{J}^{-1} \mqty( \mathcal{F}^E \\ \mathcal{F}^{L_z} ) \equiv \mqty( \dot{p}(p,x,s_\parallel) \\ \dot{x}(p,x,s_\parallel) )\,,
\end{equation}
where the Jacobian matrix is computed from the derivatives of $E(p,x),L_z(p,x)$ for spherical orbits
\begin{equation}
    \mathbb{J} = \mqty( \partial_p E & \partial_x E \\ \partial_p L_z & \partial_x L_z )\,.
\end{equation}
Additionally, we explicitly split the evolution into the geodesic part and a spin-induced deviation as
\begin{subequations}\label{eq:evolution_p_x_linearized}
\begin{align}
    p(t) &= p_\text{g}(t) + q \chi_\parallel \delta p(t) \,, \\
    x(t) &= x_\text{g}(t) + q \chi_\parallel \delta x(t) \,.
\end{align}
\end{subequations}
This allows us to write and integrate the evolution equations for the deviations $\delta p, \delta x$ in a separate step from the inspiral at geodesic order as
\begin{equation}\label{eq:evolution_delta_r_x}
    \dv{t} \mqty( \delta p \\ \delta x ) = \mqty( \partial_p \dot{p}_\text{g} & \partial_x \dot{p}_\text{g} \\ \partial_p \dot{x}_\text{g} & \partial_x \dot{x}_\text{g} ) \mqty( \delta p \\ \delta x ) + \mqty( \delta \dot{p} \\ \delta \dot{x} )\,,
\end{equation}
where
\begin{align}
    \mqty( \delta \dot{p} \\ \delta \dot{x} ) &= -\mathbb{J}^{-1}_\text{g} \qty( \delta \mathbb{J} \mqty( \dot{p}_\text{g} \\ \dot{x}_\text{g} ) + q \mqty( \delta\mathcal{F}^E \\ \delta\mathcal{F}^{L_z} ) )\,, \\
    \delta\mathbb{J} &= \mqty( \partial_p \delta E & \partial_x \delta E \\ \partial_p \delta L_z & \partial_x \delta L_z ) \, .
\end{align}
The first term on the right-hand side of Eq. \eqref{eq:evolution_delta_r_x} contains derivatives of $\mathbb{J}_\text{g}$ and $\mathcal{F}^{E,L_z}_\text{g}$ with respect to $p$ and $x$. See Appendix \ref{app:const_dervts} for the derivatives of the constants of motion.

In the two-timescale expansion, the waveform can be expressed as
\begin{equation}
    h = \frac{\mu}{r} \sum_{lmk} A_{lmk}(t) {}_{-2}S_{lm}^{a\omega_{mk}(t)}(\theta) e^{- i \Phi_{mk}(t) + i m \phi} \ , \label{eq:waveform_mulitscale}
\end{equation}
where the amplitudes $A_{lmk}$ and the frequency $\omega_{mk}$ are given by
\begin{align}
    A_{lmk}(t) &= -\frac{2 C^+_{lmk}(p_\text{g}(t), x_\text{g}(t))}{\omega_{mk}^2(t)} \,, \\
    \omega_{mk}(t) &= \omega_{mk}(p_\text{g}(t), x_\text{g}(t)) \,.
\end{align}
These quantities slowly evolve due to the evolution of $p$ and $x$. In other words, the amplitudes are only derived from adiabatic-level quantities, and the phases are the only part that carry 1PA information relevant for parameter estimation \cite{Mathews:2025,Burke:2023lno}.

After we split the phase $\Phi_{mk}$ into the azimuthal and polar parts $\Phi_{mk}(t) = m \Phi_\phi(t) + k \Phi_z(t)$, their evolution can be expressed as
\begin{equation}\label{eq:evolution_Phi_g}
    \dv{t} \mqty( \Phi_z \\ \Phi_\phi ) = \mqty( \Omega_z(p(t),x(t),s_\parallel) \\ \Omega_\phi(p(t),x(t),s_\parallel) ) \,.
\end{equation}
We again linearize the phase evolution as
\begin{align}
    \Phi_{z}(t) &= \Phi_z^\text{g}(t) + q \chi_\parallel \delta \Phi_z(t) \,, \\
    \Phi_{\phi}(t) &= \Phi_\phi^\text{g}(t) + q \chi_\parallel \delta \Phi_\phi(t) \,, 
\end{align}
This will give us evolution equations for the linear parts 
\begin{equation}\label{eq:evolution_delta_Phi}
    \dv{t} \mqty( \delta\Phi_z \\ \delta\Phi_\phi ) = \mqty( \partial_p \Omega_{z\text{g}} & \partial_x \Omega_{z\text{g}} \\ \partial_p \Omega_{\phi\text{g}} & \partial_x \Omega_{\phi\text{g}} ) \mqty( \delta p \\ \delta x ) + \mqty( \delta \Omega_z \\ \delta \Omega_\phi ) \,,
\end{equation}
where the matrix $\partial_{p,x} \Omega_{z,\phi\text{g}}$ and the vector $\delta \Omega_{z,\phi}$ are evaluated at $p_\text{g}(t)$ and $x_\text{g}(t)$.

\section{Numerical implementation and results}\label{sec:numerical_implementation}

Having established the theoretical framework, we now turn to the numerical implementation and results. We present the calculation of the trajectories and respective energy and angular momentum fluxes, their use in the calculation of the flux-driven inspirals, and the subsequent calculation of the waveforms. Then we use the waveforms to calculate mismatches between waveforms with spinning and nonspinning secondaries. These results enable us to address the central question of the possibility of detecting secondary spins in EMRIs. 

\subsection{Trajectories}
As a first step, our code computed the geodesic trajectories and velocities, and their spin-correction in the Drummond-Hughes gauge described in Sec.~\ref{sec:general_solutions_spherical_orbits}. Except for the polar component of the fiducial trajectory, and the spin correction to the radial motion~\eqref{eq:rad_shift_parallel_DH_gauge}, all of the required expressions involved Fourier series expansions, whose coefficients were computed as described in Eq.~(61) of Ref.~\cite{Skoupy:2023}. The polar geodesic trajectories were instead computed using the analytic solution available in the \textit{KerrGeodesic} package of the \textit{Black Hole Perturbation Toolkit} \cite{KerrGeodesicsZenodo,BHPToolkit}. Armed with this solution, we computed the closed-form expression for the radial spin correction given in Eq.~\eqref{eq:rad_shift_parallel_DH_gauge}. The solutions for the geodesic motion are analytical, thus their Fourier series in the mean anomaly $w_z$ converge rapidly. As shown in Ref.~\cite{Piovano:2024}, the Fourier series for the spin trajectories and velocities also exhibit fast convergence. However, it was observed in Refs.~\cite{Kerachian:2023oiw,Skoupy:2023,Piovano:2024} that the waveform partial amplitudes and fluxes are affected by the number of terms retained in the Fourier series of the trajectories, especially for highly inclined and/or eccentric orbits. To account for this, we increased the number of Fourier modes for the trajectories and velocities using the following criterion: given the inclination $x_g$ for the orbits, we used
\begin{itemize}
    \item 64 modes if $\abs{x_g} \leq 0.15$ 
    \item 48 modes if $0.15 < \abs{x_g} \leq 0.2$
    \item 32 modes otherwise.
\end{itemize}
This algorithm could be improved to find the number of Fourier modes that minimize truncation error without compromising computational speed. For the scope of the present work, the aforementioned selection criterion was sufficient.

\subsection{Fluxes}

To calculate the fluxes $\mathcal{F}^{E,L_z}$ and their derivatives, we first calculate the geodesic parts of the amplitudes $C^{\pm\text{g}}_{lmk}$, their linear parts $\delta C^+_{lmk}$, and their derivatives $\partial_{p,x} C^{\pm\text{g}}_{lmk}$ by numerically evaluating the integrals \eqref{eq:Cpm_lmk} and \eqref{eq:deltaCpm_lmk}. The principles of this calculation are described in \cite{Skoupy:2023}. However, here we calculated the homogeneous solutions of the radial and polar Teukolsky equations $R^\pm_{lm\omega}(r)$ and ${}_{-2}S_{lm}^{a\omega}(\theta)$ and their $\omega$-derivatives using the code described in \cite{Piovano:2021}. 

The algorithm for computing the eigenvalues and eigenfunctions of the polar Teukolsky equation, and their derivatives, is unstable for higher frequencies and Kerr parameters. Thus, we increased the initial precision according to the parameters $a, \omega, l, m$. Once we obtain the homogeneous solutions and their derivatives, we continue the calculation with double precision.

We calculate the amplitudes in a loop over $l$, $m$, $k$ modes and simultaneously sum the fluxes. The summation in Eqs.~\eqref{eq:fluxes} is arranged as
\begin{align}
    \mathcal{F} &= \sum_{m} \mathcal{F}_m \,, \\
    \mathcal{F}_m &= \sum_k \mathcal{F}_{mk} \,, \\
    \mathcal{F}_{mk} &= \sum_l \mathcal{F}_{lmk} \,,
\end{align}
where convergence is checked for each of the sums as specified below, and we start with the $m=2$, $k=0$, $l=2$ mode. Thanks to the symmetries \eqref{eq:symmetries_C_S}, we can compute only the modes with $\omega_{mk} > 0$ and then multiply the flux by 2. The relative precision of the summation required in our calculations is set to $\varepsilon_\Sigma = 10^{-7}$ for the geodesic fluxes and $\varepsilon_\Sigma = 10^{-4}$ for their derivatives. 

First, the sum over $l$ starts \textbf{with increasing $l$} at $l_\text{i} = \max\{m + k, \abs{m}, 2\}$, which corresponds to the mode with the highest flux in most cases. The sum ends when the flux $\abs{\mathcal{F}_{lmk}}$ monotonically decreases below $0.1 \varepsilon_\Sigma \abs{\mathcal{F}_{l=2,m=2,k=0}}$. Then the summation is repeated with decreasing $l$ starting at $l_\text{i}-1$.

Similarly, the sum over $k$ starts at $\max\{ 0, \lceil - m \Omega_\phi/\Omega_z \rceil \}$ and ends when $\abs{\mathcal{F}_{mk}} < 0.2 \varepsilon_\Sigma \abs{\mathcal{F}_{m=2,k=0}}$.

Finally, the sum over $m$ starts at $m=2$ and ends when
\begin{equation}
    \frac{\abs{\mathcal{F}_m}}{1 - \abs{\mathcal{F}_m/\mathcal{F}_{m-1}}} < 0.5 \varepsilon_\Sigma \abs{\sum \mathcal{F}_m} \,,
\end{equation}
where $\sum \mathcal{F}_m$ is the partial sum. In this formula, while assuming the exponential convergence of the series $\mathcal{F}_m$, the left-hand side corresponds to the estimate of the remainder of the series truncated at $m$.

\subsection{Flux interpolation}
To interpolate the fluxes, we introduce an auxiliary orbital parameter $v$ that replaces $p$ and fulfills    
\begin{equation}
    p = \frac{6 M + \Delta p}{v^2 \qty(1 + v \qty(\sqrt{\frac{6 M + \Delta p}{r_\text{ISSO,g}(a,x)+\Delta p}} - 1) )^2 } \,,
\end{equation}
where $r_\text{ISSO,g}$ is the radius of the geodesic ISSO and $\Delta p$ is a constant. With this definition, $p \rightarrow \infty$ when $v \rightarrow 0$ and $p \rightarrow r_\text{ISSO,g}(a,x)+\Delta p$ when $v \rightarrow 1$. We then interpolate the fluxes on an equidistant grid in the coordinates $0.1 M \leq a \leq 0.95 M$, $0.2 \leq v \leq 1$, $0.1 \leq x \leq 0.95$. For all $a$ and $x$ values, the outer boundary $v = 0.2$ is around $p \approx 150M$. The inner boundary is given by the distance from the ISSO, which we set to $\Delta p = 0.2M$.

To accurately interpolate the forcing functions $\dot{p}_\text{g}$, $\dot{x}_\text{g}$, $\delta \dot{p}$, and $\delta \dot{x}$ from Eqs.~\eqref{eq:rdot_xdot} and \eqref{eq:evolution_delta_r_x}, we exploit their PN expansion. By subtracting the 0PN and 1PN terms, we interpolate their 1.5PN term normalized by the behavior near the ISSO. The final expressions for the forcing functions calculated from the interpolated functions $\eta^{p,x}(a,v,x)$, $\delta \eta^{p,x}(a,v,x)$ are
\begin{align}
    \dot{p}_\text{g} &= -\frac{64}{5} q \left(\frac{M}{p}\right)^3 \Bigg( 1 - \frac{743}{336} \left(\frac{M}{p}\right) \nonumber \\
    &\phantom{=} + \frac{(M/p)^{3/2}}{1 - r_\text{ISSO}/p} \eta^{p}(a,v,x) \Bigg) \,, \\
    \dot{x}_\text{g} &= q \frac{a}{M} \left(\frac{M}{p}\right)^{11/2} (1-x^2) \eta^{x}(a,v,x) \,, \\
    \delta\dot{p} &= q \left(\frac{M}{p}\right)^{9/2} \frac{1}{\qty(1 - r_\text{ISSO}/p)^2} \delta\eta^{p}(a,v,x) \,, \\
    \delta \dot{x} &= q\frac{a}{M} \left(\frac{M}{p}\right)^5 \frac{1-x^2}{1 - r_\text{ISSO}/p} \delta\eta^{x}(a,v,x) \,,
\end{align}
\textbf{where we substitute the $v=v(a,p,x)$ relation to get the final $a,p,x$ dependence of the evolution equations.}

To improve the accuracy, we employ \textbf{third order} Hermite interpolation for the geodesic functions $\eta^{p,x}$, where their derivatives with respect to $p$ and $x$ are used.%
\footnote{We do not use the $a$ derivative since it would require to calculate derivatives of $R^\pm_{lm\omega}(r)$ with respect to $a$, which we did not implement. An additional implementation of these derivatives would improve the interpolation in the $a$ direction.}
\textbf{These derivatives are calculated from the ``semianalytical'' calculation of $\partial_{p,x} \mathcal{F}^{E,L_z}$ described in Sec.~\ref{sec:flux_derivatives}.} 
Since we interpolate the functions $\eta^{p,x}(a,v,x)$, we need to calculate their derivatives with respect to $v$ and $x$, for which we employ the Jacobian between $(a,p,x)$ and $(a,v,x)$. In this Jacobian, the derivative of $r_\text{ISSO}(x)$ with respect to $x$ is needed. We use the implicit function theorem and the separatrix polynomial of Stein and Warburton \cite{Stein:2019buj} to evaluate the derivative. \textbf{The linear-in-spin parts $\delta \eta^{p,x}$ were interpolated using third order Hermite interpolation without the use of derivatives.}

\begin{table*}
    \centering
    \begin{tabular}{c|c|c||c|c|c|c}
        \multicolumn{3}{c||}{Parameters} & \multicolumn{4}{c}{Relative differences} \\\hline
        $a/M$ & $p/M$ & $x$ & $\dot{p}_\text{g}$ & $\dot{x}_\text{g}$ & $\delta \dot{p}$ & $\delta \dot{x}$ \\ \hline \hline
0.20625 & 27.204  & 0.153125 & $3.8\times 10^{-7}$ & $1.2\times 10^{-5}$ & $2.1\times 10^{-5}$ & $1.7\times 10^{-5}$ \\
0.20625 & 26.109  & 0.896875 & $1.3\times 10^{-6}$ & $1.6\times 10^{-4}$ & $6.1\times 10^{-5}$ & $4.5\times 10^{-4}$ \\
0.20625 & 7.10584 & 0.153125 & $1.2\times 10^{-5}$ & $1.0\times 10^{-4}$ & $8.2\times 10^{-6}$ & $1.1\times 10^{-4}$ \\
0.20625 & 6.56688 & 0.896875 & $5.3\times 10^{-6}$ & $4.1\times 10^{-4}$ & $3.6\times 10^{-4}$ & $6.6\times 10^{-4}$ \\
0.84375 & 25.317  & 0.153125 & $2.8\times 10^{-6}$ & $6.0\times 10^{-5}$ & $1.0\times 10^{-4}$ & $8.8\times 10^{-5}$ \\
0.84375 & 19.1526 & 0.896875 & $4.4\times 10^{-5}$ & $4.2\times 10^{-3}$ & $9.9\times 10^{-4}$ & $5.3\times 10^{-3}$ \\
0.84375 & 6.19571 & 0.153125 & $1.7\times 10^{-4}$ & $5.6\times 10^{-4}$ & $1.1\times 10^{-4}$ & $6.3\times 10^{-4}$ \\
0.84375 & 3.77638 & 0.896875 & $5.8\times 10^{-4}$ & $1.6\times 10^{-2}$ & $1.3\times 10^{-2}$ & $1.7\times 10^{-2}$ \\\hline \hline
0.7375  & 25.793 & 0.153125 & $7.4\times 10^{-9}$ & $1.8\times 10^{-7}$ & $1.5\times 10^{-6}$ & $4.9\times 10^{-6}$ \\
0.7375  & 20.9206 & 0.896875 & $9.7\times 10^{-9}$ & $9.8\times 10^{-8}$ & $3.0\times 10^{-6}$ & $9.6\times 10^{-6}$ \\
0.7375  & 6.41697 & 0.153125 & $2.1\times 10^{-7}$ & $1.0\times 10^{-6}$ & $7.0\times 10^{-6}$ & $1.5\times 10^{-5}$ \\
0.7375  & 4.39197 & 0.896875 & $1.6\times 10^{-6}$ & $3.6\times 10^{-6}$ & $7.9\times 10^{-5}$ & $9.9\times 10^{-5}$
    \end{tabular}
    \caption{Relative interpolation errors at testing points with parameters $a$, $p$, and $x$. The first 8 rows have testing points misaligned with the grid in $a,p,x$, while the last 4 rows are aligned with the grid in $a$ and misaligned in $p$ and $x$.}
    \label{tab:interpolation_error}
\end{table*}

In our interpolation scheme, we use $5 \times 17 \times 9$ grid points in the $(a, v, x)$ directions. To assess the accuracy of the interpolation, we calculated the fluxes at certain points that do not align with the interpolation grid and compared the interpolated functions $\dot{p}_\text{g}, \dot{x}_\text{g}, \delta \dot{p}, \delta \dot{x}$ with the direct calculation at these points. Relative differences between the interpolated function $f_\text{I}$ and exact function $f_\text{E}$, $\abs{1- f_\text{I}(a,p,x)/f_\text{E}(a,p,x)}$, are given in Table~\ref{tab:interpolation_error}. We can see that the relative error is between $10^{-8}$ and $10^{-2}$. However, the last 4 rows of Table \ref{tab:interpolation_error} show points that align with the grid points in $a$ and are misaligned with $p$ and $x$. Since the value of $a$ does not change during the inspiral in our model, if we restrict the values of $a$ in our calculations to the grid, the interpolation error is then lower than $4 \times 10^{-6}$ for the adiabatic order. However, for full coverage of the parameter space, more grid points would be needed in $a$ to achieve $\order{q}$ accuracy for the 0PA fluxes.

\subsection{Inspirals}

The evolution of $p_\text{g}(t)$, $x_\text{g}(t)$, $\delta p(t)$, and $\delta x(t)$ was calculated numerically in \textit{Wolfram Mathematica} using the function \texttt{NDSolve} with \texttt{PrecisionGoal} and \texttt{AccuracyGoal} set to 13. 

We consider inspirals into a massive black hole with mass $M = 10^6 M_\odot$ with two values of the mass ratio $q = 10^{-5}$ and $10^{-4}$. The duration is $t_\text{f} = 1\, \text{year}$ for both cases.

\begin{figure}[t!]
    \centering
    \includegraphics[width=\linewidth]{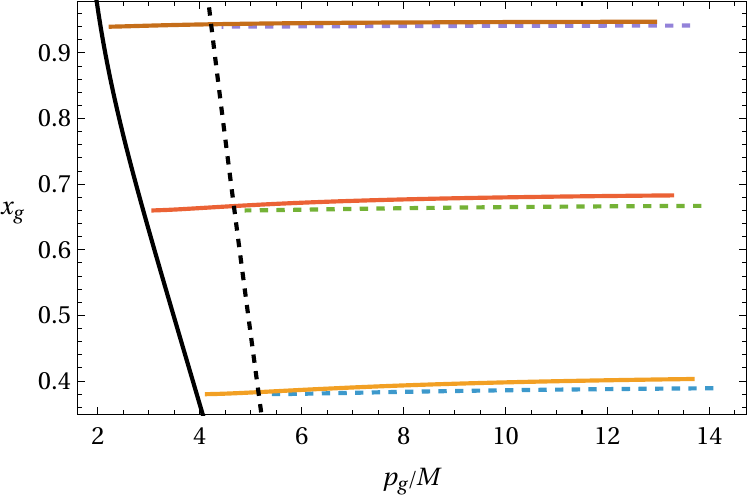}
    \caption{Evolution of the adiabatic $p_\text{g}$ and $x_\text{g}$ for $a=0.95M$ (solid) and $0.525M$ (dashed) and final $x_\text{f} = 0.38$, $0.66$, and $0.94$. The black lines show the respective ISSO positions.
    }
    \label{fig:adiabatic}
\end{figure}

The adiabatic evolution $p_\text{g}(t)$, $x_\text{g}(t)$ from Eqs.~\eqref{eq:rdot_xdot} is integrated backward in time from the endpoints at $p_\text{g}(t_\text{f}) = r_\text{ISSO} + \Delta p$ and various $x_\text{g}(t_\text{f})$ to certain values of $p_\text{g}(0)$ and $x_\text{g}(0)$. The evolution of $p_\text{g}$ and $x_\text{g}$ for $q=10^{-4}$ is shown in Fig. \ref{fig:adiabatic}. For illustration, we created an animation of the decaying orbit combined with the waveform represented as sound (see Supplemental Material \cite{SupMat}).

\begin{figure}[t!]
    \centering
    \includegraphics[width=\linewidth]{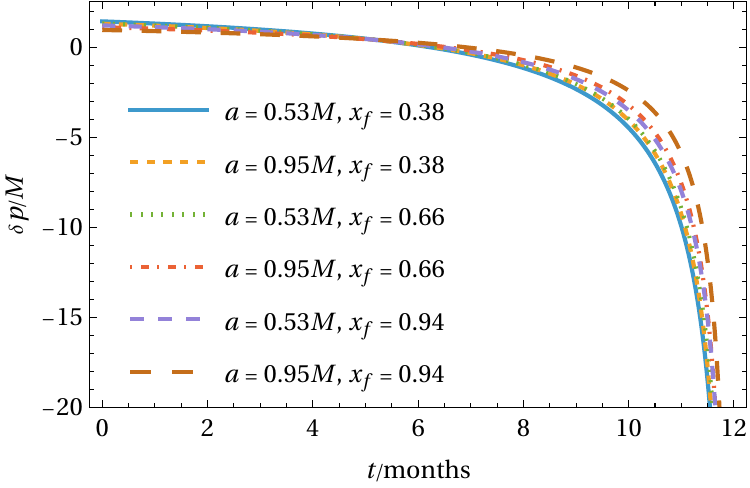}
    \includegraphics[width=\linewidth]{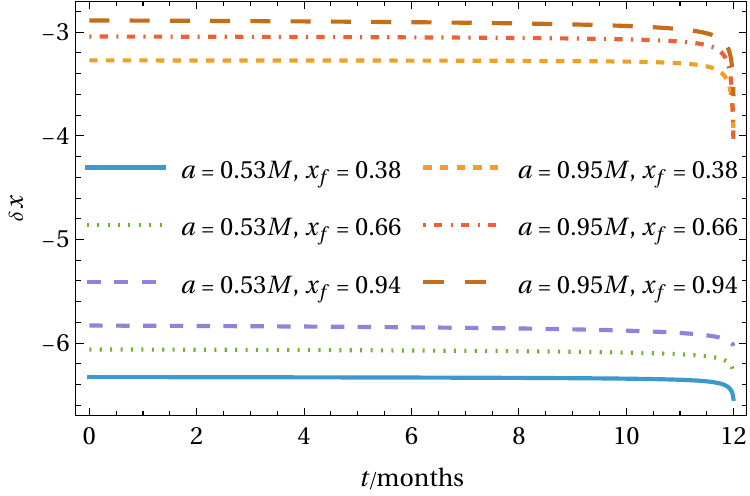}
    \caption{Evolution of the linear corrections $\delta p$ (top) and $\delta x$ (bottom) for inspirals with parameters $M=10^6 M_\odot$, $q=10^{-4}$. The inclination shifts stay roughly constant up to the ISSO, while the radius shifts dynamically evolve throughout the inspirals.}
    \label{fig:delta_r_x}
\end{figure}

Next, we integrate forward in time Eqs.~\eqref{eq:evolution_delta_r_x} for $\delta p(t)$ and $\delta x(t)$. The initial conditions are chosen such that the initial frequencies $\Omega_z(0)$ and $\Omega_\phi(0)$ are the same in the spinning and nonspinning cases, giving us the initial values
\begin{align}
    \delta p(0) &= - \frac{\partial_x \Omega_\phi \delta \Omega_z - \partial_x \Omega_z \delta \Omega_\phi}{\partial_p \Omega_z \partial_x \Omega_\phi - \partial_p \Omega_\phi \partial_x \Omega_z} \,, \\
    \delta x(0) &= \frac{-\partial_p \Omega_\phi \delta \Omega_z + \partial_p \Omega_z \delta \Omega_\phi}{\partial_p \Omega_z \partial_x \Omega_\phi - \partial_p \Omega_\phi \partial_x \Omega_z} \,.
\end{align}

\begin{figure}[t!]
    \centering
    \includegraphics[width=\linewidth]{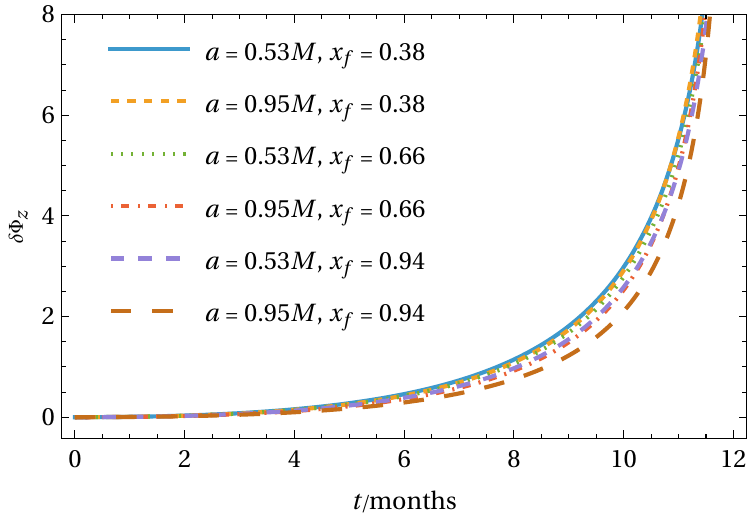}
    \includegraphics[width=\linewidth]{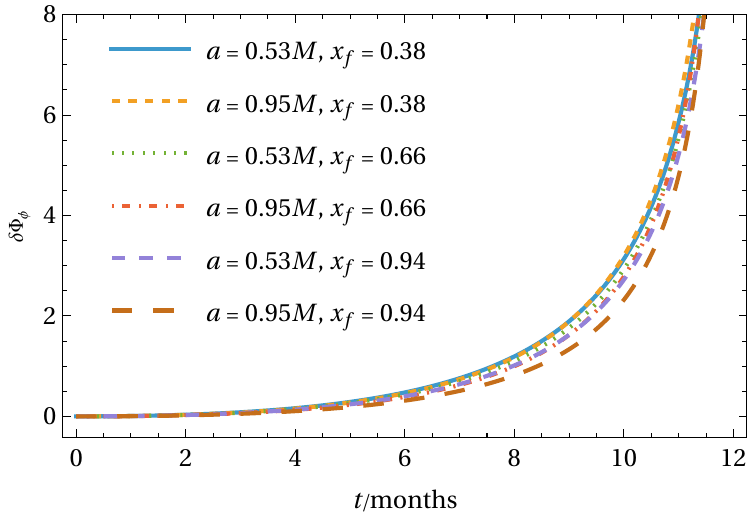}
    \caption{Evolution of the linear parts of the polar (top) and azimuthal (bottom) phases for $M=10^6 M_\odot$, $q=10^{-4}$.}
    \label{fig:delta_Phi_q1e-4}
\end{figure}

\begin{figure}[t!]
    \centering
    \includegraphics[width=\linewidth]{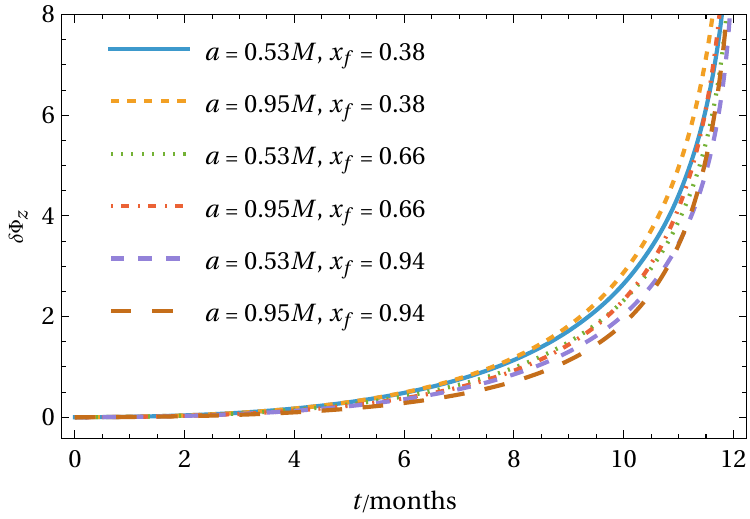}
    \includegraphics[width=\linewidth]{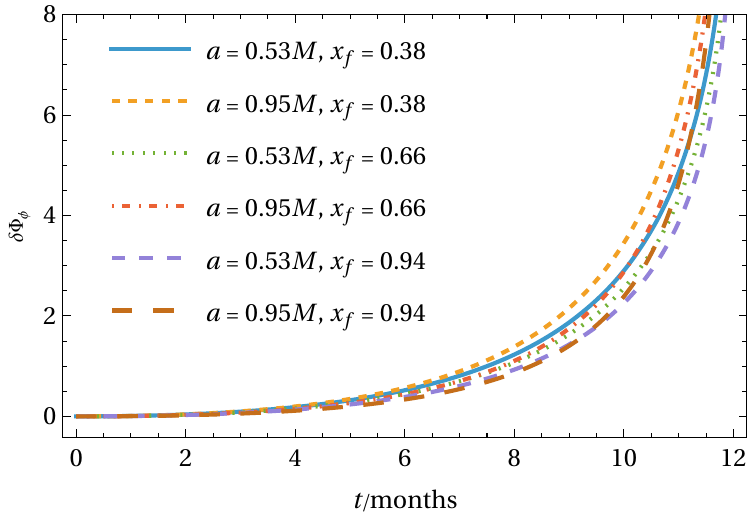}
    \caption{Evolution of the linear parts of the polar (top) and azimuthal (bottom) phases for $M=10^6 M_\odot$, $q=10^{-5}$. Interestingly, the dynamics are dominated by a universal behavior across the parameter range.}
    \label{fig:delta_Phi_q1e-5}
\end{figure}

Having obtained the evolution of the orbital parameters and their linear parts, we used them to evolve the adiabatic phases $\Phi_{z,\phi \text{g}}$ using Eqs.~\eqref{eq:evolution_Phi_g} for $s_\parallel=0$ and their linear corrections $\delta \Phi_{z,\phi}$ using Eqs.~\eqref{eq:evolution_delta_Phi}.

In Fig. \ref{fig:delta_r_x} we show the evolution of the linear corrections to the orbital parameters $\delta p(t)$ and $\delta x(t)$ for $q = 10^{-4}$. We can see that they \textbf{rapidly grow and possibly} diverge near the ISSO. This \textbf{was also observed for eccentric equatorial inspirals near the separatrix \cite{Skoupy:2022adh} and causes the linearization in spin to break down.} It is caused by the degeneracy of the various orbital relations near the ISSO, \textbf{namely the Jacobian between $(p,x)$ and $(E,L_z)$ and its derivatives diverging as $(p - r_\text{ISSO})^{-1}$ and $(p - r_\text{ISSO})^{-2}$, respectively,} and the breakdown of the two-timescale approximation \textbf{that justifies} the adiabatic radiation-reaction prescription\footnote{\textbf{This breaking point of the adiabatic prescription is, however, closer to ISSO than our inspiral endpoints.}}. 
Specifically, this also causes divergences of the functions $\delta \dot{p}$ and $\delta \dot{x}$.

\textbf{The reason for these divergencies is, for example, that the definitions of $p,e$ assume the existence of two radial turning points. However, the innermost stable orbits delineate the part of orbital space where the inner radial turning point vanishes and this leads to divergences in the various functional relations. 
However, such issues are not restricted to the $p,e,x$ parametrization since, for example, energy and angular momentum as functions of orbital configurations have local minima at the last stable orbits. This means they would also generally fail when used in parametrizations of the orbital states near these configurations.} 

\textbf{Physically, the divergences near the last stable orbits can be understood as follows. Imagine a test particle that is at the geodesic ISSO. Now add a secondary spin to the test particle that perturbs the orbit and causes it to plunge into the central black hole on a dynamical time-scale, or, in roughly one orbital period. The phase difference between the original geodesic ISSO and the plunging orbit will generally be $\mathcal{O}(1)$ over a single orbital period, while our formalism assumes that the phase difference can only be $\mathcal{O}(\chi q)$. The inability of the expansion to capture this effect manifests as the aforementioned divergence. This physical insight also suggests ways of fixing the divergence; for example, a part of the divergences could be  removed by a different spin gauge where 
stable bound orbits are mapped exclusively to stable bound orbits~\cite{Skoupy:2024uan,Piovano:2025aro}. Another part could be removed by not linearizing the evolution $p(t), x(t)$ in Eq.~\eqref{eq:evolution_p_x_linearized}.} 

In Figs. \ref{fig:delta_Phi_q1e-4} and \ref{fig:delta_Phi_q1e-5} we plot the linear corrections to the phases $\delta \Phi_{z,\phi}$ which are equivalent to the dephasing for fully aligned maximally spinning secondary, $\chi_\parallel = 1$. Similarly to $\delta p$ and $\delta x$, they \textbf{seem to be} diverging near the ISSO. Because \textbf{it is difficult to determine which part of the divergence is physical}, it is challenging to determine the final dephasing. The dephasing for both mass ratios and all primary spins and final inclinations stays below 1 radian in the first two thirds of the inspiral.

\subsection{Frequency domain waveforms}\label{sec:FD_waveforms}

Statistical analysis of GW signals is typically performed in the frequency domain. We used the extended stationary phase approximation (SPA)~\cite{Hughes:2021} to obtain an analytic approximation of the Fourier transform of our waveform.
As a first step, we rewrite Eq.~\eqref{eq:waveform_mulitscale} by taking advantage of  identities~\eqref{eq:symmetries_C_S}, 
\begin{multline}
    h(t) = \frac{G M q}{c^2 D_{\mathrm L}} \sum_{\substack{m,k \\ \omega_{mk}>0}} \Big( H_{mk}(\theta) e^{-i\Phi_{mk} + i m \phi} \\ + (-1)^k \overline{H}_{mk}(\pi-\theta) e^{i\Phi_{mk} - i m \phi} \Big) \ ,
\end{multline}
where $D_{\mathrm L}$ is the luminosity distance of the observer with respect to the inspiral and
\begin{align}\label{eq:H_mk(theta)}
    H_{mk}(\theta) &= \sum_{l=l_\text{min}}^\infty A_{lmk} \,{}_{-2}S_{lm}^{a\omega_{mn}}(\theta) \,.
\end{align}
We include all modes for which $|A_{lmk}| \geq 30^{-1} \max_{lmk}{|A_{lmk}|}$, for a total of 48 modes. We choose this value to include a representative number of the dominant modes while keeping the computational cost in check, since our implementation in \textit{Mathematica} has not been optimized for performance. Our analysis is not meant to be exhaustive, but only to qualitatively capture the importance of the spin of the smaller body. A more in-depth study is required in order to evaluate the appropriate number of modes~\cite{Khalvati:2024tzz,Speri:2023jte,Chua:2020stf}. We interpolate the combination of amplitude and spin-weighted spheroidal harmonics \eqref{eq:H_mk(theta)} for each combination of $m$ and $k$ in our selection of modes and for five different viewing angles symmetric around the equatorial plane. Before the interpolation, the functions are normalized by their behavior for large $p$ and by $(1-x^2)^\abs{k}$. Because the requirements for the accuracy of the amplitudes are lower than for the fluxes, we do not use their derivatives with respect to $p$ and $x$.

\begin{figure}
    \centering
   \includegraphics[width=\linewidth]{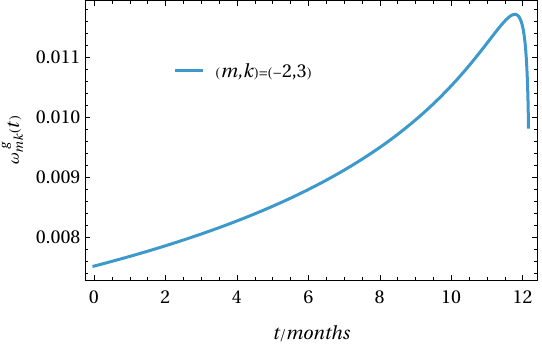}
    \caption{Example of a non-monotonic harmonic for a quasi-spherical inspiral. The plot shows the waveform frequency for the mode $(m,k)=(-2,3)$, with parameters $a = 0.9M$, $x_f =0.75$. Due to the appearance of such non-monotonous harmonics one needs to introduce the extended stationary-phase approximation.}
    \label{fig:example_non_monotonic_mode}
\end{figure}

The two polarizations $h_+$ and $h_\times$ are given by 
\begin{multline} \label{eq:hplus_SPA_monotonic}
    h_+(t) = \frac{G M q}{c^2 D_{\mathrm L}} \sum_{\substack{m,k \\ \omega_{mk}>0}} \Big( \mathcal H^+_{mk} \cos \!\big(\Phi_{mk} - \xi^+_{mk}\big) \\ +  \mathcal H^-_{mk} \cos\!\big(\Phi_{mk} - \xi^-_{mk}\big) \Big) \ ,
\end{multline}
\begin{multline}
    h_\times(t) = \frac{G M q}{c^2 D_{\mathrm L}} \sum_{\substack{m,k \\ \omega_{mk}>0}} \Big( \mathcal H^+_{mk} \sin \!\big(\Phi_{mk} - \xi^+_{mk}\big) \\ -  \mathcal H^-_{mk} \sin\!\big(\Phi_{mk} - \xi^-_{mk}\big) \Big) \ ,
\end{multline}
where
\begin{align} \label{eq:hcross_SPA_monotonic}
\mathcal H^+_{mk} &= \sqrt{(\Re H_{mk}(\theta))^2 + (\Im H_{mk}(\theta))^2} \ , \\
\mathcal H^-_{mk} &= \sqrt{(\Re \overline H_{mk}(\pi - \theta))^2 + (\Im \overline H_{mk}(\pi - \theta))^2} \ , \\
\xi^+_{mk} &= m\phi + \arg H_{mk}(\theta) \ ,\\
\xi^-_{mk} &= m\phi - \arg \Big((-1)^k \overline H_{mk}(\pi - \theta) \Big) \ .
\end{align}
Here $\arg \kappa =\arctan(\Im \kappa/\Re \kappa)$ is the argument of the complex number $\kappa$.
For all the modes $(m,k)$ for which $\Phi_{mk}$ is monotonic, we implemented the standard SPA, which reads
\begin{align}
\tilde h_+(f) &= \sum_{\substack{m,k \\ \omega_{mk}>0}} e^{i \Psi_{mk}}\sqrt{\frac{2\pi}{\lvert\dot\omega^{\rm g}_{mk}\rvert}}\mathcal A_{mk} \ , \\
\tilde h_{\times}(f) &= \sum_{\substack{m,k \\ \omega_{mk}>0}} ie^{i \Psi_{mk}}\sqrt{\frac{2\pi}{\lvert\dot\omega^{\rm g}_{mk}\rvert}} \mathcal A_{mk} \ ,
\end{align}
where $f=\omega/(2\pi)$, and
\begin{align}
\mathcal A_{mk} &= \frac{1}{2} \bigg(\mathcal H^+_{mk}  e^{-i\xi^+_{mk}} + \mathcal H^-_{mk} e^{-i \xi^-_{mk}} \bigg) \ , \\
\Psi_{mk} &= 2\pi f \tilde t_{\rm g}(f) - \Phi_{mk}(t_{\rm g}(f)) - \frac{\pi}{4} \text{sgn}(\dot \omega_{mk}) \ ,\\
\dot \omega^\text{g}_{mk} &= m\ddot \Phi_\phi^\text{g}(t) + k\ddot \Phi_z^\text{g}(t) \ .
\end{align}
The waveform polarizations~\eqref{eq:hplus_SPA_monotonic} and~\eqref{eq:hcross_SPA_monotonic} are evaluated at the SPA time $t=\tilde t_{\rm g}(f)$, which obey the following relation
\begin{equation}
 2 \pi f = \omega^{\rm g}_{mk}(\tilde t_{\rm g}(f)) \ , \label{eq:root_t_g_SPA}
\end{equation}
For monotonic modes, solving the previous non-linear equation is equivalent to solving the following ODE
\begin{equation}
\dv{\tilde t_{\rm g}}{f} = \frac{2\pi}{\dot \omega^{\rm g}_{mk}(\tilde t_{\rm g}(f))}   \ . \label{eq:root_t_g_SPA_ODE}
\end{equation}
Thus, the SPA time $\tilde t_{\rm g}(f)$ depends implicitly on the harmonic mode $(m,k)$. Moreover, we can show the SPA phase $\Psi_{mk}$ only depends on the leading order SPA time $\tilde t_{\rm g}(f)$ up to $\mathcal O(q^2 \chi^2)$. Indeed, we can write
\begin{widetext}
\begin{align}
\begin{split}
&2\pi f \Big(\tilde t_{\rm g}(f) +q \chi \delta \tilde t(f) \Big) - \Phi^{\rm g}_{mk}(t_{\rm g}(f)) - q \chi \delta \Phi_{mk}(t_{\rm g}(f)) - q \chi \dot \Phi^{\rm g}_{mk}(t_{\rm g}(f)) \delta \tilde t(f) - \frac{\pi}{4} \text{sgn}(\dot \omega_{mk}) + \mathcal O(q^2 \chi^2) \\
&= \Psi_{mk} +  q \chi \delta \tilde t(f) \Big(  2\pi f  - \omega^{\rm g}_{mk}(t_{\rm g}(f)) \Big) + \mathcal O(q^2 \chi^2) \ ,
\end{split}
\end{align}
\end{widetext}
and the term $  2\pi  f  - \omega^{\rm g}_{mk}(t_{\rm g}(f))$ is zero thanks to Eq.~\eqref{eq:root_t_g_SPA}.
EMRI waveforms for quasi-spherical orbits can also have modes $(m,k)$ for which $\omega_{mk}$ is no longer monotonic, but it reaches a maximum, and then decreases (see Fig.~\ref{fig:example_non_monotonic_mode}). Such modes require the implementation of the extended SPA~\cite{Hughes:2021,Speri:2023jte}, which reads
 \begin{align}
\tilde h_+(f) &= \displaystyle \sum^{2}_{v=1}  \sum_{\substack{m,k \\ \omega^{(v)}_{mk}>0}} e^{i \Psi^{(v)}_{mk}} \mathcal B^{(v)}_{mk}\mathcal A^{(v)}_{mk} \ , \label{eq:hplus_SPA_extended} \\
\tilde h_{\times}(f) &= \sum^{2}_{v=1}  \sum_{\substack{m,k \\ \omega^{(v)}_{mk}>0}}ie^{i \Psi^{(v)}_{mk}}\mathcal B^{(v)}_{mk}\mathcal A^{(v)}_{mk} \ , \label{eq:hcross_SPA_extended}
\end{align}
with
\begin{align}
  \mathcal B^{(v)}_{mk} &= \frac{i\dot\omega^{\text{g}(v)}_{mk}}{\lvert\ddot \omega^{\text{g}(v)}_{mk}\rvert} K_{1/3}\big(i X^{(v)}_{mk}\big)e^{i X^{(v)}_{mk}} \ , \\
  X^{(v)}_{mk} &= - \frac{1}{3}\frac{(\dot\omega^{\text{g}(v)}_{mk})^3}{(\ddot\omega^{\text{g}(v)}_{mk})^2} \ ,
\end{align}
and $ K_n(z)$ is the modified Bessel function of the second kind. The index $(v)$ in the previous expressions refers to the two branches of the roots of Eq.~\eqref{eq:root_t_g_SPA}. For the non-monotonic modes for quasi-spherical inspirals, $\dot \omega^{\rm g}_{mk}>0$ in the first branch $\tilde t^{(1)}_{\rm g}(f)$, whereas $\dot \omega^{\rm g}_{mk}\leq 0$ in the second branch $\tilde t^{(2)}_{\rm g}(f)$. We solved Eq.~\eqref{eq:root_t_g_SPA_ODE} to obtain $\tilde t^{(1)}_{\rm g}(f)$, while we found it more convenient to get $\tilde t^{(2)}_{\rm g}(f)$ as the root of Eq.~\eqref{eq:root_t_g_SPA} using the \texttt{FindRoot} routine of \textit{Mathematica}.

We tested the validity of the extended SPA against frequency domain waveforms obtained with the fast Fourier transform (FFT). There are a few technical details to consider when comparing the two approximations of the Fourier transform. A detailed comparison is presented in Appendix~\ref{app:comparison_waveforms}. We found that the extended SPA is in excellent agreement with the FFT, as observed in~\cite{Hughes:2021,Speri:2023jte, Piovano:2022ojl}. Moreover, the SPA waveforms are well suited for \textbf{quick forecasts on the statistical errors of the parameters, at least for conservative SNRs
$\sim 30$ (as discussed in Appendix~\ref{app:comparison_waveforms}). However, there are many uncertainties in EMRI formation rates, and therefore expected SNRs, which may be of order $\mathcal O(100)$. For such high SNRs, the SPA may introduce relevant systematic errors. An analysis of these possible systematics is left for future work, as it is outside of the scope of this paper.}

\begin{figure}
    \centering
    \includegraphics[width=\linewidth]{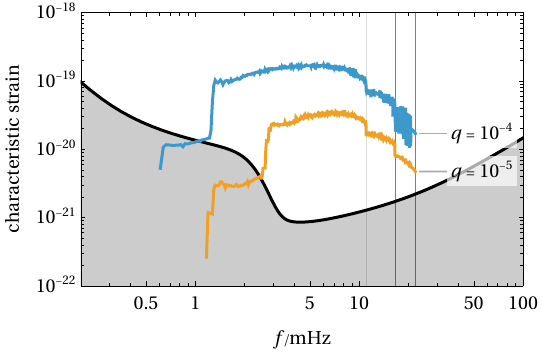}
    \caption{
    Characteristic strain $2 f \sqrt{\abs{\tilde{h}_+(f)}^2 + \abs{\tilde{h}_\times(f)}^2}$ for inspirals with $M = 10^6 M_\odot$, $a = 0.95M$, $\chi_\parallel = 0$, $x_\text{f} = 0.66$, $\theta = \pi/4$, $\phi = 0$, $D_{\mathrm L} = 1\, {\rm Gpc}$  plotted against the LISA noise curve $\sqrt{f S_n(f)}$ (black). \textbf{Vertical lines show frequency $f = m \Omega_{\phi}(a,r_{\rm{ISSO}}+0.2M,x_{\rm f})/(2\pi)$ for $m = 2,3,4$.}
    }
    \label{fig:characteristic_strain}
\end{figure}

\subsection{Comparisons between waveforms for a spinning and non-spinning particle}\label{sec:comaprisons_waveforms}

We are now ready to estimate the impact of the secondary spin by calculating the mismatches between waveforms for a spinning and non-spinning particle. For this purpose, we first compute the \textbf{total} overlap between two time-domain waveforms $p = p_+ - i p_\times$ and $q = q_+ - i q_\times$, which is defined as
\begin{equation}
    \mathcal O(p,q) = \frac{(p | q)}{\sqrt{(p | p)(q | q)}}  \ , \label{eq:overlap}
\end{equation}
\textbf{where} the noise-weighted scalar product between two waveforms $p$ and $q$ is given by
\begin{equation}\label{eq:waveforms_product}
   (p|q) =  4 \Re \int^\infty_0 \dd f\frac{\overline{\tilde p(f)} \tilde q(f)}{S_n(f)} \ ,
\end{equation}
$S_n(f)$ is the one-sided noise spectral density of the LISA detector~\cite{Robson:2018ifk}, which includes the confusion noise from the unresolved Galactic binaries, while the overline denotes complex conjugation. Waveforms comparison are typically performed using the optimized \textbf{total} overlap over the initial time $t_\text{i}$ and phase $\phi_\text{i}$, also known as \textbf{total} faithfulness (see for instance~\cite{Nitz:2013mxa,Chatziioannou:2017tdw})
\begin{equation}
  \mathcal F(p,q) = \text{max}_{t_\text{i},\phi_\text{i}}  \mathcal O(p,q)\,.
\end{equation}
The \textbf{total} mismatch (or \textbf{total} unfaithfulness) between two waveforms is
\begin{equation}
    \mathcal{M}( h^{s \neq 0}, h^{s=0} ) \equiv 1 - \mathcal{F}( h^{s \neq 0}, h^{s=0} ) \,.
\end{equation}

In our study, we consider a primary with mass $M = 10^6 M_\odot$, two mass ratios $q = 10^{-5}, 10^{-4}$, and vary the secondary and primary spin $\chi_\parallel$ and $a$, and final inclination $x_\text{f}$. The viewing angles are $\theta = \pi/4, \phi = 0$ and the initial phase of the polar motion is chosen such that the secondary body starts at $z(0) = \sqrt{1-x^2}$. For illustration, we show the frequency domain waveforms against the LISA noise curve in Fig.~\ref{fig:characteristic_strain}.

\begin{figure}[tb]
    \centering
    \includegraphics[width=\linewidth]{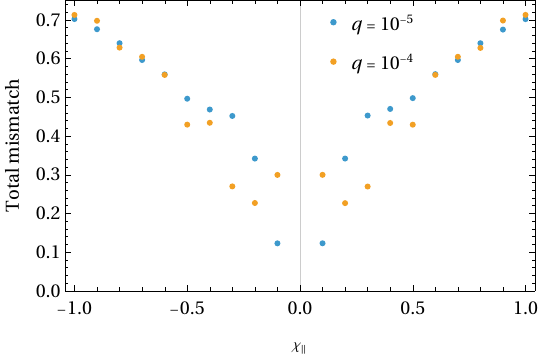}
    \caption{Dependence of the mismatch optimized over relative phase between waveforms (unfaithfulness) of inspirals with spinning and nonspinning secondary on the smaller body spin. The primary spin and final inclination are fixed to $a=0.95M$, $x_\text{f} = 0.66$.}
    \label{fig:mismatches_faith_spin}
\end{figure}

\begin{figure}[tb]
    \centering
    \includegraphics[width=\linewidth]{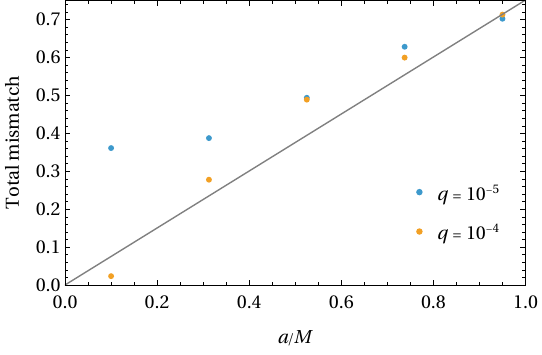}
    \caption{Dependence of the mismatch optimized over relative phase between waveforms (unfaithfulness) of inspirals with spinning and nonspinning secondary on the primary spin. The final inclination and secondary spin are fixed to $x_\text{f} = 0.66$, $\chi_\parallel = 1$. \textbf{The gray line shows the dependence $(a/M) \sqrt{1-x_{\rm f}^2}$.}}
    \label{fig:mismatches_faith_a}
\end{figure}

\begin{figure}[tb]
    \centering
    \includegraphics[width=\linewidth]{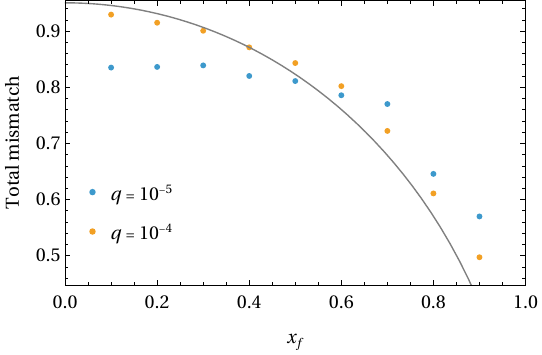}
    \caption{Dependence of the mismatch optimized over relative phase between waveforms (unfaithfulness) of inspirals with spinning and nonspinning secondary on the final inclination. The spins are fixed to $a=0.95M$, $\chi_\parallel = 1$. \textbf{The gray line shows the dependence $(a/M) \sqrt{1-x_{\rm f}^2}$.}}
    \label{fig:mismatches_faith_x}
\end{figure}

Figures~\ref{fig:mismatches_faith_spin} and \ref{fig:mismatches_faith_a} present the unfaithfulness for several values of secondary and primary spin of the binary, respectively, while Fig. \ref{fig:mismatches_faith_x} show how the unfaithfulness evolves with the final orbital inclination $x_\text{f}$ \footnote{Note that the inclination parameter $x$ is 1 (-1) for prograde (retrogade) equatorial orbits and $|x|$ decreases as the orbit approaches polar orbit.}. A common trend can be observed in all plots: the mismatches are mostly independent of the mass ratios. There are some exceptions, like for $a=0.1$ in Fig.~\ref{fig:mismatches_faith_a} and $x_f \leq 0.4$ in Fig.~\ref{fig:mismatches_faith_x}. \textbf{However, the $a=0$ limit should be taken more carefully since here we match the initial $\Omega_\phi$ and $\Omega_z$ which are not independent for $a=0$. We leave this endeavor for future work.} 

In general, the unfaithfulness increases with the primary spin and the absolute value of the secondary, while it decreases with the final orbital inclination. Figures~\ref{fig:mismatches_faith_a} and~\ref{fig:mismatches_faith_x} suggest that the detectability of secondary spin will be lower for nearly equatorial inspirals or a slowly spinning primary black hole, which is in agreement with previous studies~\cite{Huerta:2011kt,Huerta:2011zi,Piovano:2021,Burke:2023lno}. The mismatch for fully aligned, maximally spinning secondary at the viewing angle $\theta = \pi/4$ can be approximately expressed as $\mathcal{M} \approx (a/M) \sqrt{1 - x_\text{f}^2}$.

According to the Lindblom criterion~\cite{Lindblom:2008cm}, two waveforms are distinguishable if their mismatch $\mathcal M \geq \frac{{\cal D}}{{2\rm SNR}^2}$, with ${\cal D}$ the number of parameters of the waveform model. If we neglect the spin precession of the smaller body, we have in total twelve parameters, since we assumed that the primary spin is aligned to the $z$-axis. \footnote{It is worth noticing that the precession of the binary spins and the orbital plane are not independent degrees of freedom. In the large mass ratio limit, a quasi-spherical EMRI can be viewed as an asymmetric binary with precessing spins on quasi-circular orbits. The latter orbital configuration has the same degrees of freedom as a spherical inspiral.} For ${\cal D} =12$, the mismatch threshold is $\mathcal M \simeq  6 \times 10^{-3}$ for a prototypical SNR=30. Thus, waveforms including the secondary spin are distinguishable from waveforms with a non-spinning body across all of the parameter space. However, these mismatches provide no information on the probability distribution underlying the secondary spin. We discuss these aspects in the next Section. 

\subsection{Discussion of mismatch results} 

As already noted, the Lindblom criterion \cite{Lindblom:2008cm} provides a sufficient condition for \textbf{distinguishability of two waveforms} by computing waveform mismatches. For typical LISA EMRIs, the Lindblom criterion is satisfied in practically all configurations presented in Figs.~\ref{fig:mismatches_faith_spin}--\ref{fig:mismatches_faith_x}, apart from the case of very small primary spins $a$. In particular, the result $\mathcal{M} \approx (a/M) \sqrt{1 - x_\text{f}^2}$ suggests that the detectability of secondary spin can be interpreted as particularly boosted thanks to ``spin-spin'' effects. 

Does this mean that LISA data analysis will definitely be able to extract secondary spins in spherical EMRI waveforms? Not necessarily. \textbf{The issue is that in a complete parameter estimate, even a discernible change to the waveform can be absorbed by a small variation of the other parameters of the waveform within its uncertainty range. In other words, the large mismatches do not necessarily imply that the small body spin is detectable, since we have not investigated whether the waveform variations caused by secondary spin have sizable correlations with the other parameters.\footnote{The ``effectualness" between waveforms~\cite{Thompson:2025hhc} may provide more insights on the link between the observed mismatches and possible systematic biases in the parameters. We leave this investigation for future work.}} \citet{Burke:2023lno} demonstrated that the Lindblom criterion is indeed not sufficient for the detectability of any given effect. They showed that large mismatches between waveforms for quasi-circular equatorial inspirals, including and neglecting the secondary spin, did not translate into detectability, even for relatively high SNRs. This is because spin effects in quasi-circular inspirals can be effectively mimicked by a variation of other parameters of the EMRI (see also Refs~\cite{Huerta:2011kt,Huerta:2011zi,Piovano:2021}).   

What, then, can we say about such degeneracies for spherical rather than quasi-circular inspirals? Let us picture a short stretch of a waveform generated by a quasi-circular inspiral contrasted with that generated by a generic quasi-spherical inspiral. Short stretches of quasi-circular inspirals are characterized by a single frequency $\Omega_\phi$ and a drift $\dot{\Omega}_\phi$, while quasi-spherical inspirals are characterized by four parameters $\Omega_\phi, \dot{\Omega}_\phi,\Omega_\theta, \dot{\Omega}_\theta$ at any instance. While a small shift in the primary parameters $M,a$ can absorb a variation of $\Omega_\phi,\dot{\Omega}_\phi$ at any given referential point (say at the moment of peak strain in the waveform), this is generally no longer possible for the four parameters $\Omega_\phi, \dot{\Omega}_\phi,\Omega_\theta, \dot{\Omega}_\theta$. Similar arguments lead to an expectation that the richer harmonic content of spherical inspirals will be deformed by the secondary spin in a way that cannot be easily mimicked by shifts of other parameters, thus making it easier to measure than in quasi-circular inspirals. 

This expectation is currently supported by the preliminary Fisher matrix analyses with kludge waveforms presented by \citet{Cui:2025bgu}. An important observation in their work is that the orbital inclination breaks the statistical correlations between the parameters observed in quasi-circular inspirals in Refs.~\cite{Huerta:2011kt,Huerta:2011zi,Piovano:2021,Burke:2023lno}. This being said, let us also note that the study of~\citet{Cui:2025bgu} has a few caveats. Their analysis relies on Fisher matrix estimates, which are notoriously ill-conditioned~\cite{Vallisneri:2007ev,Burke:2020vvk,Gair:2012nm,Porter:2015eha,Piovano:2021} (see Ref.~\cite{Kejriwal:2025upp} on how to mitigate the numerical instability). Moreover, a Fisher matrix forecast relies on  the assumption that the underlying posterior distribution of the parameters can be approximated, for large SNR, as a Gaussian distribution. Ref.~\cite{Burke:2023lno} shows that the posterior distribution is, in fact, non-Gaussian when the secondary spin is included as a parameter.
While it considered generic orbits, Ref.~\cite{Cui:2025bgu} employed kludge waveforms based on PN gravitational fluxes and the quadrupolar approximation for the amplitudes, and it only includes the conservative effects due to the secondary spin. These approximations may have a sizable impact on the detectability of secondary spin, since kludge waveforms can have significant systematic errors compared to their fully relativistic counterparts~\cite{Katz:2021yft,Khalvati:2024tzz,Burke:2023lno,Chapman-Bird:2025xtd}.

In conclusion, on one hand, our results bring about renewed optimism about the detectability of secondary spin by LISA. On the other hand, only a careful Bayesian data analysis with relativistic waveforms can fully assess which bounds can be put on secondary spin in EMRIs.

\section{Summary and conclusions} \label{sec:Concl}
This work provides the first fully relativistic treatment of secondary spin effects in nearly-spherical EMRI waveforms, establishing both the computational framework and the preliminary evidence that these effects will be detectable by LISA. The results suggest that LISA will be able to measure not just the masses and primary spins of EMRI systems, but also the spins of the smaller companions—opening a new window into the formation and evolution of compact objects in galactic nuclei.

Our work can also be seen as the first exploration of the detectability of the subleading spin parameters in GW inspirals of fully precessing spinning binaries at large mass ratios in a more general context. As such, it complements similar studies for inspirals of comparable-mass-ratio binaries in the LIGO-Virgo-KAGRA band such as Refs \cite{Vitale:2014mka,Schmidt:2014iyl,Purrer:2015nkh, Gerosa:2020aiw}.

We started by specializing the Hamilton-Jacobi formalism from Ref.~\cite{Piovano:2024} to nearly spherical orbits. By working in both (almost) fixed constants of motion and fixed turning points gauges, we found analytical expressions for linear-in-spin parts of the frequencies and constants of motion for nearly spherical orbits (see Supplemental Material \cite{SupMat}). Then we derived numerical expressions for the Fourier expansions of the trajectory.

In the next step, we used the orbits to find respective energy and angular momentum fluxes. We derived expressions for the numerical calculation of the geodesic and linear-in-spin parts of the fluxes, along with their derivatives with respect to the orbital parameters. Subsequently, we calculated the fluxes on a grid in the parameter space and interpolated them using Hermite interpolation. We found that the interpolation error of both the adiabatic and linear-in-spin fluxes is sufficiently small in the majority of the parameter space\textbf{, at least when considering fixed values of primary spin $a/M$. For interpolation also over primary spin, the interpolation strategy would likely provide sufficient accuracy for the linear-in-spin corrections to the fluxes. However, more refined strategies will be needed for the leading adiabatic terms.}

Next, we showed that spherical orbits remain spherical even for inspirals with a spinning secondary. Then we computed the evolution of the orbital parameters in the adiabatic and linear-in-spin order. Subsequently, we calculated frequency domain waveforms through SPA. We found that in the quasi-spherical case, the frequency of certain modes is nonmonotonous, which was previously observed only for eccentric equatorial inspirals. Thus, we implemented extended SPA to tackle this problem.

Finally, we calculated the unfaithfulness between inspirals with a spinning and nonspinning secondary at various points in the parameter space. Figs.~\ref{fig:mismatches_faith_spin}--\ref{fig:mismatches_faith_x} show that the unfaithfulness is mostly independent of mass ratio, while it clearly depends on the primary spin and inclination. For instance, the unfaithfulness is $\sim 0.75$ for highly inclined orbits, but it decreases with the inclination, dropping to $\sim 0.3$ for almost equatorial configurations. A similar trend is observed for the binary spins. For highly spinning primary and secondary spins, the unfaithfulness is $~0.75$, but it is significantly lower for a slowly spinning binary, reaching $~0.1$ for a slowly spinning configuration. 

Even such large mismatches as those found here are not entirely sufficient to state that secondary spin will necessarily be observable by LISA. This is because parametric waveform degeneracies were already observed to effectively absorb spin effects of similar magnitude in the case of quasi-circular inspirals. However, it is plausible that the richer and more complex structure of spherical-inspiral waveforms breaks the degeneracies and that secondary spin will be detectable in this case.

This work serves mostly as a proof-of-concept, not as an exhaustive test of the secondary spin detectability. For this, a full Bayesian analysis with Fisher matrices or a Markov chain Monte Carlo analysis would need to be carried out, which could reveal degeneracies in the parameter space (see \cite{Piovano:2021,Burke:2023lno}). Before that, however, we would have to optimize the speed of the evaluation of the waveforms as it was done, e.g., in the \textit{FastEMRIWaveforms} framework \cite{Chua:2020stf,Katz:2021yft,Chapman-Bird:2025xtd}, since such computations require many evaluations. We leave this analysis to future work.

This work contributes to the generation of phase-accurate, fully relativistic gravitational waveforms from EMRIs by providing all the spin-proportional 1PA terms in the phasing for quasi-spherical inspirals. It can be extended by generalizing the motion to fully generic bound orbits, where the recent results from Ref.~\cite{Skoupy:2024uan} can be utilized. Additionally, other 1PA effects such as conservative self-force could be included in the model similarly to Ref. \cite{Lynch:2024}. Another challenging piece would be the inclusion of second-order gravitational fluxes, which are currently known only for the quasi-circular case and a slowly spinning primary \cite{Warburton:2021kwk}. 

For demonstrational and educational purposes, we created a video in which we combined an animation of the inspiraling orbit with the waveform represented as sound (see Supplemental Material \cite{SupMat}).

\begin{acknowledgments}

V.S. and V.W. are supported by the Charles U. \textit{Primus} Research Program 23/SCI/017.
GAP was supported by an Irish Research Council Fellowship under grant number GOIPD/2022/496 during part of this work. G.A.P. also acknowledges the support of the Win4Project grant ETLOG of the Walloon Region for the Einstein Telescope. 
We would like to thank Lisa Drummond and Philip Lynch for reading an earlier version of this paper and giving us helpful feedback and the referee for many useful comments and suggestions. 
This work makes use of the \textit{Black Hole Perturbation Toolkit} \cite{BHPToolkit}, specifically the \textit{KerrGeodesics} \cite{KerrGeodesicsZenodo} package and a custom-made version of the \textit{SpinWeightedSpheroidalHarmonics} package, version 0.3.0~\cite{BHPToolkitSpinWeightedSpheroidalHarmonicsv0.3.0}.

\end{acknowledgments}

The data that support the findings of this article are openly available \cite{DataZenodo,CodeGitHub}.

\appendix

\section{Shift to the innermost stable spherical orbit}\label{app:rISSO}

As described by \citet{Stein:2019buj} (see also \cite{Levin:2008yp}), the radius of the innermost stable spherical geodesic in Kerr space-time at a generic $x_{\rm g}\in(-1,1)$ is expressed by finding the roots of a polynomial of order 12. We use the \textit{Mathematica} implementation of the root-finding algorithm in the \textit{KerrGeodesics} package~\cite{KerrGeodesicsZenodo} to find high-accuracy geodesic values of the ISSO radius, energy, and angular momentum. The shift to the ISSO is then computed with the help of the recently presented analytical solution for spinning-particle orbits \cite{Skoupy:2024uan} as follows.

In Ref. \cite{Skoupy:2024uan} the final orbital solution is expressed as\footnote{Note that we add a tilde over any shift related to the virtual trajectory here as compared to Ref. \cite{Skoupy:2024uan}. This is meant to avoid confusion with quantities such as $\delta r$ in the main text.}
\begin{align}
    t(\tilde{\lambda}; C) &= t_\text{g}(\tilde{\lambda}; \Tilde{C}) - \frac{3 s_\parallel}{2 \sqrt{K_{\rm g}}} \tau_\text{g}(\tilde{\lambda}; \Tilde{C}) \nonumber \\ 
    &\phantom{=} - \delta \tilde{t}(r_\text{g}(\tilde{\lambda}),z_\text{g}(\tilde{\lambda}),\psi(\tilde{\lambda})) \, , \\
    x^k(\tilde{\lambda}; C) &= x^k_\text{g}(\tilde{\lambda}; \Tilde{C}) - \delta \tilde{x}^k(r_\text{g}(\tilde{\lambda}),z_\text{g}(\tilde{\lambda}),\psi(\tilde{\lambda})) \, , \label{eq:xkanalytic} \\
    \tau(\tilde{\lambda}; C) &= \qty( 1 - \frac{3 s_\parallel E_{\rm g}}{2 \sqrt{K_{\rm g}}} ) \tau_\text{g}(\tilde{\lambda}; \Tilde{C}) \, , 
\end{align}
where $\tilde{\lambda}$ is a deformed Mino parameter, $\tilde{C} = \tilde{E},\tilde{L},\tilde{K}$ is a certain virtual set of constants of motion, and we denote collectively $x^k = r,z,\phi$. The functions $\tau_{\rm g}(\_,\_),t_{\rm g}(\_,\_),x^k_{\rm g}(\_,\_)$ are functions corresponding to the analytical solution for the corresponding quantities along a geodesic with the given constants of motion parametrized by Mino time. Finally, $\delta \tilde{t}, \delta \tilde{x}^k$ are additional shifts of the worldline given in Ref.~\cite{Skoupy:2024uan}. The important property is that the shifts are $\mathcal{O}(\chi q)$ and only a function of the position of the particle and the constants of motion; they do not grow with time and are bounded along any trajectory. 

The virtual constants are related to the constants of the spinning-particle motion as 
\begin{align}
    & \tilde{E} = E + \frac{s_\parallel(1 - E^2)}{2\sqrt{K}} \,, \label{eq:Etilde} \\
    & \tilde{L} = L_z + \frac{s_\parallel(a - L_z E/2)}{\sqrt{K}} \,, \\
    & \tilde{K} = K + \frac{s_\parallel( 3 a (L_z-aE) - K E )}{\sqrt{K}} \,. \label{eq:Ktilde}
\end{align}
We can now find the last spherical orbit as follows. The boundedness of the function $r_{\rm g}(\_,\_)$ appearing in the $x^k = r$ component of Eq.~\eqref{eq:xkanalytic} determines whether the radial motion is bounded or unbounded, since the $\delta r$ term is always bounded. We can then find ``virtual-spherical'' orbits by requiring that $r_{\rm g}(\tilde{\lambda},\tilde{C}) =$ const. By finding the marginally stable orbits within this class, we also indirectly identify the ISSO corresponding to the construction of spherical orbits in this paper. This is because the virtual-spherical orbits will correspond to spherical orbits with an $\mathcal{O}(\chi q)$ orbital eccentricity. However, such small linear-order shifts in eccentricity do not affect the constants of motion at leading order near spherical orbits because of the degeneracy of the radial potential for spherical orbits. 

In the main text, we parametrize the orbit by $E_{\rm g},L_{z\rm g},K_{\rm g}$. These can be identified by using the $\mathtt{KerrGeoISSO[a,\tilde{x}]}$ function to find the radius $\tilde{r}$ of the virtual-geodesic ISSO with inclination parameter $\tilde{x}$, the function $\mathtt{KerrGeoConstantsOfMotion[a,\tilde{r},0,\tilde{x}]}$ then returns the $\tilde{C}$ constants, which are then transformed to $E_{\rm g},L_{z\rm g},K_{\rm g}$ according to equations \eqref{eq:Etilde} to \eqref{eq:Ktilde}.

To transform to orbital elements defined by averaged turning points, we still need to average the $\delta \tilde{r}$ and $\delta \tilde{z}$ terms at turning points. These contain parts that oscillate with the precession phase $\psi$ and average out to zero, and parts that are independent on the precession phase, which read
\begin{align}
    & \delta \tilde{r}_{\parallel} = s_\parallel \frac{r_{\rm ISSO,g} \left((r^2_{\rm ISSO,g} + a^2) E_{\rm g} - a L_{z\rm g} \right)}{\sqrt{K_{\rm g}} (r^2_{\rm ISSO,g} + a^2 z^2_{\rm g})} \,, \\
    & \delta \tilde{z}_{\parallel} = s_\parallel \frac{a z_{\rm g} \left(L_{z\rm g} - a E(1-z^2_{\rm g}) \right)}{\sqrt{K_{\rm g}}(r^2_{\rm ISSO,g} + a^2 z^2_{\rm g})} \,.
\end{align}
The radius is constant to leading order, so we only need to average the $\delta \tilde{r}$ shift over $z_{\rm g}$. The $\delta \tilde{z}_\parallel$ shift then only needs to be evaluated at the turning points and requires no further averaging. This yields
\begin{align}
     x(\tilde x) =&  \tilde{x} - s_\parallel \frac{a(1 - x^2_{\rm g}) \left(L_{z\rm g} - a E_{\rm g} x^2_{\rm g} \right)}{\tilde{x} \sqrt{K_{\rm g}} \left(r^2_{\rm ISSO,g} + a^2 (1-\tilde{x}^2)\right)} \,, 
     \\
    r(\tilde{x}) = & r_{\rm ISSO, g}(\tilde{x}) - s_\parallel \frac{(r^2_{\rm ISSO,g} + a ^2) E_{\rm g} - a L_{z\rm g}}{\sqrt{K_{\rm g}}} \frac{\mathsf{\Pi}(\gamma_r | k_z)}{\mathsf{K}(k_z)} \,,
\end{align}
where $\mathsf{K}$ and $\mathsf{\Pi}$ are elliptic integrals of the first and third kind, respectively, \textbf{and $k_z$ and $\gamma_r$ are defined in Eq.~\eqref{eq:kz_gammar}}. This also allows us to perform consistency checks between the formalisms. In Fig.~\ref{fig:rISSO}, we parametrize the ISSO by the actual $x$ instead of $\tilde{x}$. For that purpose, we have to add a transformation term and finally obtain the shift of the original trajectory expressed as Eqs.~\eqref{eq:rISSO} and \eqref{eq:rISSOshift} in the main text.

\section{Spherical inspirals through the virtual-trajectory formalism}\label{app:vtInsp}
In the virtual-trajectory formalism in Ref. \cite{Skoupy:2024uan} the motion is described using a shifted or ``virtual'' trajectory $\tilde{x}^\mu = x^\mu + \delta \tilde{x}^\mu$ where $\delta \tilde{x}^\mu \sim \mathcal{O}(q\chi)$ at all times and for all trajectories (including near-spherical trajectories). The advantage of this shift is that the equation became separable on the shifted worldline. In particular, one obtains relations such as
\begin{align}
    & \tilde{K} = - \Delta(\tilde{r}) \tilde{u}_r^2 + \mathcal{H}(\tilde{r},\tilde{E},\tilde L_z)\,, \label{eq:tKr}\\
    & \mathcal{H} = \frac{1}{\Delta (\tilde{r})} \left[\tilde{E}(\tilde{r}^2 + a^2) - a \tilde L_z \right]^2 - \tilde{r}^2
\end{align}
where $\tilde{u}_r$ is the covariant $r$-component of $\dd \Tilde{x}^\mu / \dd \tau$, and the tilde constants of motion were defined in Eqs.~\eqref{eq:Etilde}--\eqref{eq:Ktilde}. The important point here is that this radial relation is functionally identical to the radial motion of Kerr geodesics.

We would now like to answer the question: if the \textit{virtual} trajectory is exactly spherical, $\tilde{r} = {\rm const.}, \tilde{u}_r = 0, \dot{\tilde{u}}_r = 0$, will it remain spherical under radiation reaction? To answer this question, we will resort to a variant of the argument given by  Kennefick and Ori for geodesics \cite{Kennefick:1995za}.

Consider first the condition for a (virtual) exactly spherical orbit to stay spherical under radiation-reaction. The sphericity of the orbit requires that the following conditions are met
\begin{align}
    &-\tilde{K} + \mathcal{H}(\tilde{r},\tilde{E},\tilde L_z) =0\,,\\
    & \frac{\partial}{\partial \tilde{r}}\left(-\tilde{K} + \mathcal{H}(\tilde{r},\tilde{E},\tilde L_z) \right) = 0 \,. \label{eq:sphcond}
\end{align}
If these relations have zero time derivatives during radiation-reaction, then one can show that the following relation must be fulfilled
\begin{align}
    \frac{\dd \tilde{K}}{\dd \tau}\Big|_{\rm spher.} = \frac{\partial \mathcal{H}}{\partial \tilde{E}} \frac{\dd \tilde{E}}{\dd \tau} + \frac{\partial \mathcal{H}}{\partial \tilde L_z} \frac{\dd \tilde L_z}{\dd \tau}\,.
\end{align}
We can now take a time derivative of relation \eqref{eq:tKr} and using $\tilde{u}_r = 0$ and Eq. \eqref{eq:sphcond} we see it is always fulfilled. In the second half of Ref. \cite{Kennefick:1995za}, Kennefick and Ori investigated the question of whether this solution of the equations of the evolution of the Carter constant can be understood as stable and physical. Consider a quantity
\begin{align}
    \tilde{K}_{\rm ns} = \tilde{K}_{\rm spher.}(\tilde{E},\tilde L_z) + \Delta(\tilde{r}) \tilde{u}_r^2 - \mathcal{H}(\tilde{r},\tilde{E},\tilde L_z) \,,
\end{align}
where $\tilde{K}_{\rm spher.}(\tilde{E},\tilde L_z)$ is the Carter constant for exactly spherical orbits expressed as a function of $\tilde{E},\tilde L_z$. Obviously, $\tilde{K}_{\rm ns}=0$ for spherical orbits and nonzero away from spherical orbits. By close analogy with Kennefick and Ori we can then expand the expression for $\dd \tilde{K}_{\rm ns}/\dd \tau$ to quadratic order in deviations from spherical orbits and show that the time-derivative will develop forced oscillations due to the local self-force. However, since the forcing itself will only be a functional on the original spherical orbit, it will only be oscillating with harmonics of the polar frequency $\tilde{\omega}_z$. On the other hand, the radial oscillations will oscillate with a characteristic oscillation period $\tilde{\omega}_r$. 
As a result, the time-average of $\dd \tilde{K}_{\rm ns}/\dd \tau$ evaluates to zero unless $2 k \tilde{\omega}_z - \tilde{\omega}_r =0$ for some $k \in \mathbb{Z}$, which is not the case along spherical orbits in Kerr space-time. As a result, there is no secular growth of ``non-sphericity'' for the virtual trajectory.

Since the original MPD trajectory is always $\mathcal{O}(s)$ close to the virtual trajectory by $x^\mu = \tilde{x}^\mu - \delta x^\mu$, the original MPD trajectory corresponding to a spherical trajectory has at most $\mathcal{O}(q\chi)$ eccentricity. In fact, all trajectories with $\mathcal{O}(q\chi)$ eccentricities have the same set of constants of motion up to $\mathcal{O}(q^2\chi^2)$, so all of them correspond to near-spherical virtual trajectories that stay near-spherical as per the Kennefick-Ori argument above. As a result, any of our trajectories with $\mathcal{O}(q\chi)$ eccentricity can be assumed to keep an $\mathcal{O}(q\chi)$ eccentricity at all times throughout the inspiral.

\section{Derivatives of the constants of motion}\label{app:const_dervts}

For our evolution equations \eqref{eq:rdot_xdot} and \eqref{eq:evolution_delta_r_x} we need up to second order derivatives of the geodesic energy $E_\text{g}$ and angular momentum $L_{z\text{g}}$ with respect to $p$ and $x$ and first order derivatives of the linear correction $\delta E$ and $\delta L_z$. In the fixed turning points gauge these derivatives can be obtained from conditions at the turning points in Eqs. \eqref{eq:conditions_turning_points_geodesic} and \eqref{eq:condition_turning_points_linear}. By taking derivatives of these equations with respect to $p$ or $z_{1 \rm g} = \sqrt{1-x_{\rm g}^2}$ we can derive the relations for the derivatives of the constants of motion
\begin{align}
    \pdv{p} \mqty( E_{\rm g} \\ L_{z \rm g} \\ K_{\rm g} ) &= - \mathbb{M}^{-1} \mqty( R_\text{g}' \\ R_\text{g}'' \\ 0 ) \,, \\
    \pdv{z_{1\rm g}} \mqty( E_{\rm g} \\ L_{z \rm g} \\ K_{\rm g} ) &= - \mathbb{M}^{-1} \mqty( 0 \\ 0 \\  Z'_\text{g} ) \,,
\end{align}
\begin{widetext}
\begin{align}
    \pdv{p}  \mqty( \delta E \\ \delta L_z \\ \delta K ) &= - \mathbb{M}^{-1} \qty( \pdv{\vec{v}_\text{s}}{p} + \qty( \pdv{\mathbb{M}}{p} + \pdv{\mathbb{M}}{E_{\rm g}} \pdv{E_{\rm g}}{p} + \pdv{\mathbb{M}}{L_{z\rm g}} \pdv{L_{z\rm g}}{p} ) \mqty( \delta E \\ \delta L_z \\ \delta K ) ) \,, \\
    \pdv{z_1}  \mqty( \delta E \\ \delta L_z \\ \delta K ) &= - \mathbb{M}^{-1} \qty( \pdv{\vec{v}_\text{s}}{z_{1\rm g}} + \qty( \pdv{\mathbb{M}}{z_{1\rm g}} + \pdv{\mathbb{M}}{E_{\rm g}} \pdv{E_{\rm g}}{z_{1\rm g}} + \pdv{\mathbb{M}}{L_{z\rm g}} \pdv{L_{z\rm g}}{z_{1\rm g}} ) \mqty( \delta E \\ \delta L_z \\ \delta K ) ) \,,
\end{align}
\begin{align}
    \pdv[2]{p}  \mqty( E_{\rm g} \\ L_{z\rm g} \\ K_{\rm g} ) &= - \mathbb{M}^{-1} \qty( \mqty( R_\text{g}'' \\ R_\text{g}''' \\ 0 ) + \qty( 2 \pdv{\mathbb{M}}{p} + \pdv{\mathbb{M}}{E_{\rm g}} \pdv{E_{\rm g}}{p} + \pdv{\mathbb{M}}{L_{z\rm g}} \pdv{L_{z\rm g}}{p} ) \mqty( \partial_{p} E_{\rm g} \\ \partial_{p} L_{z\rm g} \\ \partial_{p} K_{\rm g} ) ) \,, \\
    \pdv[2]{z_{1\rm g}}  \mqty( E_{\rm g} \\ L_{z \rm g} \\ K_{\rm g} ) &= - \mathbb{M}^{-1} \qty( \mqty( 0 \\ 0 \\ Z_\text{g}'' ) + \qty( 2 \pdv{\mathbb{M}}{z_{1\rm g}} + \pdv{\mathbb{M}}{E_{\rm g}} \pdv{E_{\rm g}}{z_{1\rm g}} + \pdv{\mathbb{M}}{L_{z\rm g}} \pdv{L_{z\rm g}}{z_{1\rm g}} ) \mqty( \partial_{z_{1\rm g}} E_{\rm g} \\ \partial_{z_{1\rm g}} L_{z\rm g} \\ \partial_{z_{1\rm g}} K_{\rm g} ) ) \,, \\
    \frac{\partial^2}{\partial p \partial z_{1\rm g}} \mqty( E_{\rm g} \\ L_{z\rm g} \\ K_{\rm g} ) &= - \mathbb{M}^{-1} \qty( \qty( \pdv{\mathbb{M}}{z_{1\rm g}} + \pdv{\mathbb{M}}{E_{\rm g}} \pdv{E_{\rm g}}{z_{1\rm g}} + \pdv{\mathbb{M}}{L_{z\rm g}} \pdv{L_{z\rm g}}{z_{1\rm g}} ) \mqty( \partial_{p} E_{\rm g} \\ \partial_{p} L_{z\rm g} \\ \partial_{p} K_{\rm g} ) + \pdv{\mathbb{M}}{p} \mqty( \partial_{z_{1\rm g}} E_{\rm g} \\ \partial_{z_{1\rm g}} L_{z \rm g} \\ \partial_{z_1\rm g} K_{\rm g} ) ) \,.
\end{align}
The matrix $\mathbb{M}$ is independent of $K_{\rm g}$. Here, a prime over the geodesic radial (polar) potential $R_{\rm g}$ ($Z_{\rm g}$) denotes a derivative with respect to to $r_{\rm g}$ ($z_{\rm g}$), and similar notation applies for high order derivatives.
\end{widetext}

\section{Validating SPA waveforms} \label{app:comparison_waveforms}
In this Section, we assessed the validity of the SPA against frequency domain waveform obtained with the FFT of a time-domain signal. The main tool in our analysis is the overlap  defined in~\eqref{eq:overlap}. We considered two inspiral configurations: an almost equatorial inspiral ending with $x_\text{f} = 0.93$ and an almost polar inspiral with $x_\text{f} = 0.1$. In both cases, the binary has spins $a=0.95$, $\chi=1$, masses $M=10^6 M_\odot$, $\mu = 10 M_\odot$, while the inspiral ends at $p = r_{\text{ISSO}} + 0.2$, with $r_{\text{ISSO}}$ the location of the geodesic ISSO. For both configurations, we set the viewing angles as $(\theta,\phi)=(\pi/4,0)$. The results are presented in Table~\ref{tab:overlap_FFT_vs_SPA}.

There are a few technical points to take into account when comparing the two approximations of the Fourier Transform. First of all, a time domain waveform must be sampled appropriately when employing a discrete Fourier Transform to avoid a distortion called aliasing. Following the Nyquist-Shannon theorem, we sampled the time-domain waveform with a time interval $\Delta t_{\text{sample}} > 1/(2f_{\text{max}})$, with $f_{\text{max}}$ given by
\begin{equation}
   f_{\text{max}} = \frac{1}{2\pi} \text{max}_{(m,k)}\big( \omega^{\rm g}_{mk}(T) \big)
\end{equation}
with $T = 1 \text{ year}$. We found that the overlap $\mathcal O(h^{\text{SPA}},h^{\text{FFT}})$ is sensitive to the number of samples $N = T/\Delta t_{\text{sample}}$ adopted to discretize the time-domain waveform. Typically, a smaller $\Delta t_{\text{sample}}$ leads to an FFT with larger $\mathcal O(h^{\text{SPA}}_\alpha,h^{\text{FFT}}_\alpha)$ than an FFT obtained using fewer samples. 
Thus, we increase the number of samples by reducing  $\Delta t_{\text{sample}}$ until the overlap  $\mathcal O(h^{\text{SPA}},h^{\text{FFT}})$ remains constant, up to numerical error.

Moreover, the Fourier Transform is formally defined for functions with support over $\mathbb R$. In practical applications, signals have a finite time length, which has an impact in the frequency domain. On one hand, any discrete Fourier transform suffers, to a certain degree, from spectral leakage, which makes it more difficult to resolve the signal's spectrum. On the other hand, for each $(m,k)$ mode, an SPA template is zero outside the frequency support $[\omega_{mk}(0), \omega_{mk}(T)]$. In other words, an SPA waveform starts and ends abruptly in the spectrum. To mitigate spectral leakage, we employed a Tukey window, whose effect can be adjusted by tuning a parameter $\beta \in [0,1]$. When $\beta =0$ ($\beta =1$), the Tukey window reduces to the rectangular (Hann) window. We tapered the time-domain signal with the Tukey window before performing the FFT for different values of $\beta$. To obtain the equivalent effect in the frequency domain, we convolve the SPA waveform with the FFT of the Tukey window~\cite{Speri:2023jte}.  We noticed that the overlap $\mathcal O(h^{\text{SPA}}_\alpha,h^{\text{FFT}}_\alpha)$ remains practically unchanged for any $\beta \leq 10^{-2}$ for a sufficiently small $\Delta t_{\text{sample}}$, whereas a window with $\beta \geq 10^{-2}$ led to smaller overlaps.

\begin{table}[tb!]
\centering
\begin{tabular}{cc}
\hline
\hline
$x_{\rm f}$ & $\mathcal O(h^{\text{SPA}},h^{\text{FFT}})$\\
\hline 
0.1 & 0.9987 \\
\hline
0.93 & 0.9972 \\
\hline
\hline
\end{tabular}
\caption{\textbf{Total} overlap $\mathcal O(h^{\rm{SPA}},h^{\rm{FFT}})$ 
between frequency-domain waveforms obtained with the SPA, and by performing the FFT to the time domain signal, for two different values of the ending inclination $x_\text{f}$. 
In both cases, we considered masses $M=10^6 M_\odot$, $\mu = 10 M_\odot$,  spins $a=0.95$, $\chi=1$. The EMRIs evolve for one year up $p = r_{\text{ISSO}} + 0.2$, with $r_{\text{ISSO}}$ the location of the geodesic ISSO.}
\label{tab:overlap_FFT_vs_SPA}
\end{table}

Table~\ref{tab:overlap_FFT_vs_SPA} shows that the SPA waveform model matches relatively well with the FFT waveform: $\mathscr{F}(h^{\rm{SPA}},h^{\rm{FFT}})\gtrsim 0.997$, in agreement with~\cite{Hughes:2021,Speri:2023jte}. According to the Lindblom criterion~\cite{Lindblom:2008cm}, two waveforms are distinguishable if their overlap $\mathcal O $ is smaller than $\mathcal O\sim 1- \frac{{\cal D}}{{2\rm SNR}^2}$, with ${\cal D}$ the number of parameters of the waveform model. In our case, ${\cal D} =12$, and assuming a prototypical \textbf{conservative} $\text{SNR} =30$, we get $\mathcal O \simeq 0.993$. Thus, the SPA is sufficiently accurate for \textbf{EMRI binaries with lower SNRs}, as observed in~\cite{Speri:2023jte, Chapman-Bird:2025xtd}. 
\textbf{For sources with sufficiently high SNR, the SPA may introduce observable systematic errors in the waveform. However, as mentioned in the main text, the Lindblom criterion alone is insufficient to assess the importance of waveform systematics. For instance, Ref.~\cite{Piovano:2022ojl} compared the statistical uncertainties recovered with FFT and SPA waveforms in the special case of spins-aligned, circular, equatorial orbits. The mismatches between SPA and FFT waveforms seen in~\cite{Piovano:2022ojl} are similar to the ones shown in Table II here. However, the statistical uncertainties recovered with Fisher matrices using FFT and SPA waveforms were in perfect agreement with each other, with differences at the level of few percent.}

\textbf{A key feature missing in this work is the 
detector response, which in the case of LISA is given by highly nontrivial numerical functions with complicate time and frequency dependence. As such, the detector response is difficult to incorporate in a SPA waveform. In fact, the LISA response has to be first transformed in the frequency domain, and then convolve it with the frequency domain waveform. Such a procedure may be computationally too expensive in realistic data analysis studies. However, it is still possible to directly fold the detector response in the SPA in the low-frequency regime using the long-wavelength approximation~\cite{Cutler:1997ta,Barack:2003fp}, as shown in~\cite{Piovano:2022ojl}.}

\textbf{In conclusion, frequency domain waveforms in conjunction with Fisher matrix methods can be useful to perform computationally cheap forecasts on the constraints of the binary parameters. However, SPA waveforms may not be practical or reliable for more realistic data analysis studies, and may introduce systematic errors relevant for bright sources.}

\FloatBarrier
\bibliography{paper}

\end{document}